\documentclass[zpreprint,zbstdefault]{./LaTeX/zeus/zeus_paper}
\usepackage[english]{babel}
\chardef\usc=95
\chardef\til=126
\catcode`\@=11 
\DeclareRobustCommand\xdotspace{\futurelet\@let@token\@xdotspace}
\def\@xdotspace{%
  \ifx\@let@token.\else
  \ifx\@let@token\bgroup.\else
  \ifx\@let@token\egroup.\else
  \ifx\@let@token\/.\else
  \ifx\@let@token\ .\else
  \ifx\@let@token~.\else
  \ifx\@let@token!.\else
  \ifx\@let@token,.\else
  \ifx\@let@token:.\else
  \ifx\@let@token;.\else
  \ifx\@let@token?.\else
  \ifx\@let@token/.\else
  \ifx\@let@token'.\else
  \ifx\@let@token).\else
  \ifx\@let@token-.\else
  \ifx\@let@token\@xobeysp.\else
  \ifx\@let@token\space.\else
  \ifx\@let@token\@sptoken.\else
   .\space
   \fi\fi\fi\fi\fi\fi\fi\fi\fi\fi\fi\fi\fi\fi\fi\fi\fi\fi}
\catcode`\@=12 

\newcommand{\stru}[2]{%
   \relax\ifmmode\hbox{\vrule height#1 depth#2 width0pt}%
   \else\vrule height#1 depth#2 width0pt\fi}

\newcommand{\Ronum}[1]{\uppercase\expandafter{\romannumeral#1}}
\newcommand{\ronum}[1]{\expandafter{\romannumeral#1}}
\DeclareRobustCommand{\LaTeXZ}{%
  \LaTeX\kern-.05em4\kern-.1em
  {\raisebox{-0.2ex}{$\scriptstyle\text{ZEUS}$}}\xspace}

\DeclareMathAlphabet{\mathbf}{OT1}{cmr}{bx}{sl}
\newcommand{\eVdist}{\kern-0.06667em}

\newcommand{\slashfrac}[2]{%
  \raisebox{0.5ex}{\ensuremath #1}\kern-0.12em/\kern-0.08em
  \raisebox{-.8ex}{\ensuremath #2}}

\newcommand{\sqr}[3]{%
    {\vcenter{\hrule height.#3ex\hbox{\vrule width.#2ex height#1ex
     \kern#1ex\vrule width.#3ex}\hrule height.#2ex}}}

\catcode`\@=11 
\newcommand{\parenbar}{\mathpalette\p@renb@r}
\def\p@renb@r#1#2{\vbox{%
  \ifx#1\scriptscriptstyle \dimen@.7em\dimen@ii.2em\else
  \ifx#1\scriptstyle \dimen@.8em\dimen@ii.25em\else
  \dimen@1em\dimen@ii.4em\fi\fi \offinterlineskip
  \ialign{\hfill##\hfill\cr
    \vbox{\hrule width\dimen@ii}\cr
    \noalign{\vskip-.3ex}%
    \hbox to\dimen@{$\mathchar300\hfil\mathchar301$}\cr
    \noalign{\vskip-.3ex}%
    $#1#2$\cr}}}
\catcode`\@=12 

\newcommand{\IP}{{\rm I$\kern-0.01667em$P}\xspace}

\mathchardef\qsm=63
\mathchardef\pls=43
\mathchardef\mns=512
\mathchardef\plm=518
\mathchardef\eql=61
\mathchardef\smallleft=300
\mathchardef\smallright=301
\mathchardef\les=316
\mathchardef\gre=318
\mathchardef\leq=532
\mathchardef\grq=533

\catcode`\@=11 
\newcounter{pict@width}
\newcounter{pict@height}
\newlength{\pict@scale}
\newcommand{\psfigadd}[4]{%
\setcounter{pict@width}{1*\ratio{#2+\pict@scale/2}{\pict@scale}}
\setcounter{pict@height}{1*\ratio{#3+\pict@scale/2}{\pict@scale}}
\setlength{\unitlength}{\pict@scale}
\hbox to #2{\hspace{-\fill}\begin{picture}(\thepict@width,\thepict@height)
\put(0,0){\psfig{figure=#1,width=#2,height=#3,clip=}}
\SetScale{0.283466457}
\SetWidth{1.763889}
{#4}
\end{picture}}
}
\newcounter{pict@widthfst}
\newcounter{pict@widthscd}
\newcounter{pict@widthtot}
\newcommand{\psfigaddtwo}[7]{%
\setcounter{pict@widthfst}{1*\ratio{#2+\pict@scale/2}{\pict@scale}}
\setcounter{pict@widthscd}{1*\ratio{#2+#4+\pict@scale/2}{\pict@scale}}
\setcounter{pict@widthtot}{1*\ratio{#2+#4+#6+\pict@scale/2}{\pict@scale}}
\setcounter{pict@height}{1*\ratio{#3+\pict@scale/2}{\pict@scale}}
\setlength{\unitlength}{\pict@scale}
\hbox{\hspace{-\fill}\begin{picture}(\thepict@widthtot,\thepict@height)
\put(0,0){\psfig{figure=#1,width=#2,height=#3,clip=}}
\put(\thepict@widthscd,0){\psfig{figure=#5,width=#6,height=#3,clip=}}
\SetScale{0.283466457}
\SetWidth{1.763889}
{#7}
\end{picture}}
}
\newcommand{\psfigror}[4]{%
\setcounter{pict@width}{1*\ratio{#2+\pict@scale/2}{\pict@scale}}
\setcounter{pict@height}{1*\ratio{#3+\pict@scale/2}{\pict@scale}}
\setlength{\unitlength}{\pict@scale}
\hbox{\begin{picture}(\thepict@width,\thepict@height)
\put(0,\thepict@height){\psfig{figure=#1,width=#3,height=#2,clip=,angle=270}}
\SetScale{0.283466457}
\SetWidth{1.763889}
{#4}
\end{picture}}
}
\newcommand{\psfigrol}[4]{%
\setcounter{pict@width}{1*\ratio{#2+\pict@scale/2}{\pict@scale}}
\setcounter{pict@height}{1*\ratio{#3+\pict@scale/2}{\pict@scale}}
\setlength{\unitlength}{\pict@scale}
\hbox{\begin{picture}(\thepict@width,\thepict@height)
\put(0,0){\psfig{figure=#1,width=#3,height=#2,clip=,angle=90}}
\SetScale{0.283466457}
\SetWidth{1.763889}
{#4}
\end{picture}}
}
\catcode`\@=12 
\newlength\listtextwidth

\catcode`\@=11 
\newlength{\@tabfninsert}
\newlength{\@tabfnwidth}
\newcommand{\tabfootnote}[2]{%
  \setlength{\@tabfninsert}{0.8em}
  \setlength{\@tabfnwidth}{\textwidth}
  \addtolength{\@tabfnwidth}{-\@tabfninsert}
  \addtolength{\@tabfnwidth}{-0.4em}
  \noindent\makebox[\@tabfninsert][r]{\footnotesize$^{#1}$\hfil}\hfill%
  \parbox[t]{\@tabfnwidth}{\footnotesize #2\hfill}}
\catcode`\@=12 

\begin{document}
\title{
\vspace{-5cm}
\begin{flushright} {\normalsize \tt DESY 08-201}\\ \vspace{-.25cm}{\normalsize 
\tt December 2008} \end{flushright}
\vspace{2cm}
Measurement of \boldmath $D^{\pm}$ and $D^{0}$  \unboldmath production in deep inelastic scattering using a lifetime tag at HERA
}
                    
\author{ZEUS Collaboration}
\draftversion{6.1}
\date{\today}

\abstract{
The production of $D^{\pm}$ and $D^{0}$ mesons has been measured with the ZEUS detector at HERA using an integrated luminosity of 133.6 pb$^{-1}$. The measurements cover the kinematic range $5 < Q^{2} < 1000$ GeV$^{2}$, $0.02 < y < 0.7$, $1.5 < p_{T}^{D} < 15$ GeV and $|\eta^{D}| < 1.6$. Combinatorial background to the $D$ meson signals is reduced by using the ZEUS microvertex detector to reconstruct displaced secondary vertices.
Production cross sections are compared with the predictions of next-to-leading-order QCD which is found to describe the data well. Measurements are extrapolated to the full kinematic phase space in order to obtain the open-charm contribution, $F_{2}^{c\bar{c}}$, to the proton structure function, $F_{2}$. 
}

\makezeustitle

\def\3{\ss}                                                                                        
\pagenumbering{Roman}                                                                              
                                                   %
\begin{center}                                                                                     
{                      \Large  The ZEUS Collaboration              }                               
\end{center}                                                                                       
  S.~Chekanov,                                                                                     
  M.~Derrick,                                                                                      
  S.~Magill,                                                                                       
  B.~Musgrave,                                                                                     
  D.~Nicholass$^{   1}$,                                                                           
  \mbox{J.~Repond},                                                                                
  R.~Yoshida\\                                                                                     
 {\it Argonne National Laboratory, Argonne, Illinois 60439-4815, USA}~$^{n}$                       
\par \filbreak                                                                                     
  M.C.K.~Mattingly \\                                                                              
 {\it Andrews University, Berrien Springs, Michigan 49104-0380, USA}                               
\par \filbreak                                                                                     
  P.~Antonioli,                                                                                    
  G.~Bari,                                                                                         
  L.~Bellagamba,                                                                                   
  D.~Boscherini,                                                                                   
  A.~Bruni,                                                                                        
  G.~Bruni,                                                                                        
  F.~Cindolo,                                                                                      
  M.~Corradi,                                                                                      
\mbox{G.~Iacobucci},                                                                               
  A.~Margotti,                                                                                     
  R.~Nania,                                                                                        
  A.~Polini\\                                                                                      
  {\it INFN Bologna, Bologna, Italy}~$^{e}$
% ...
\newpage                                                                                           
\enlargethispage{5cm}                                                                              
$^{\    1}$ also affiliated with University College London,                                        
United Kingdom\\                                                                                   
$^{\    2}$ now at University of Salerno, Italy \\                                                 
% ...
$^{\    5}$ supported by the research grant no. 1 P03B 04529 (2005-2008) \\                        
% ...
%
                                                           %
                                                           %
\begin{tabular}[h]{rp{14cm}}                                                                       
$^{a}$ &  supported by the Natural Sciences and Engineering Research Council of Canada (NSERC) \\  
$^{b}$ &  supported by the German Federal Ministry for Education and Research (BMBF), under        
          contract numbers 05 HZ6PDA, 05 HZ6GUA, 05 HZ6VFA and 05 HZ4KHA\\                         
$^{c}$ &  supported in part by the MINERVA Gesellschaft f\"ur Forschung GmbH, the Israel Science   
          Foundation (grant no. 293/02-11.2) and the U.S.-Israel Binational Science Foundation \\  
$^{d}$ &  supported by the Israel Science Foundation\\                                             
$^{e}$ &  supported by the Italian National Institute for Nuclear Physics (INFN) \\                
$^{f}$ &  supported by the Japanese Ministry of Education, Culture, Sports, Science and Technology 
          (MEXT) and its grants for Scientific Research\\                                          
$^{g}$ &  supported by the Korean Ministry of Education and Korea Science and Engineering          
          Foundation\\                                                                             
$^{h}$ &  supported by the Netherlands Foundation for Research on Matter (FOM)\\                   
$^{i}$ &  supported by the Polish State Committee for Scientific Research, project no.             
          DESY/256/2006 - 154/DES/2006/03\\                                                        
$^{j}$ &  partially supported by the German Federal Ministry for Education and Research (BMBF)\\   
$^{k}$ &  supported by RF Presidential grant N 1456.2008.2 for the leading                         
          scientific schools and by the Russian Ministry of Education and Science through its      
          grant for Scientific Research on High Energy Physics\\                                   
$^{l}$ &  supported by the Spanish Ministry of Education and Science through funds provided by     
          CICYT\\                                                                                  
$^{m}$ &  supported by the Science and Technology Facilities Council, UK\\                         
$^{n}$ &  supported by the US Department of Energy\\                                               
$^{o}$ &  supported by the US National Science Foundation. Any opinion,                            
findings and conclusions or recommendations expressed in this material                             
are those of the authors and do not necessarily reflect the views of the                           
National Science Foundation.\\                                                                     
$^{p}$ &  supported by the Polish Ministry of Science and Higher Education                         
as a scientific project (2006-2008)\\                                                              
$^{q}$ &  supported by FNRS and its associated funds (IISN and FRIA) and by an Inter-University    
          Attraction Poles Programme subsidised by the Belgian Federal Science Policy Office\\     
$^{r}$ &  supported by an FRGS grant from the Malaysian government\\                               
\end{tabular}                                                                                      
                                                           %
                                                           %

\pagenumbering{arabic} 
\pagestyle{plain}
\section{Introduction}
\label{sec-int}

Charm quarks are copiously produced in deep inelastic scattering (DIS) at HERA. At sufficiently high photon virtuality, $Q^{2}$, the production of charm quarks constitutes up to $30\%$ of the $ep$ cross section \cite{epj:c12:35, pl:b528:199}. Previous measurements of $D^{*\pm}$ cross sections \cite{epj:c12:35, pl:b528:199, pl:b407:402, np:b545:21,pr:d69:012004} indicate that the production of charm quarks in DIS in the range $1 < Q^{2} < 1000$\,GeV$^{2}$ is consistent with the calculations of perturbative Quantum Chromodynamics (pQCD) in which charm is predominantly produced via boson-gluon fusion (BGF). This implies that the charm cross section is directly sensitive to the gluon density in the proton.

A charm quark in the final state can be identified by the presence of a corresponding 
charmed hadron. In this paper a study of the production of two such charmed particles, the 
$D^{\pm}$ and $D^{0}/\bar{D}^{0}$ mesons, is presented. The mesons are reconstructed using the decays 
$D^{+} \rightarrow K^{-}\pi^{+}\pi^{+}$ and $D^{0} \rightarrow K^{-}\pi^{+}$, which are chosen as both contain charged 
particles\footnote{The charge-conjugated modes are implied throughout this paper.} which are well reconstructed in the ZEUS detector. The proper decay lengths are of the order 300 ${\rm \mu}$m and \mbox{100 ${\rm \mu}$m} for the $D^{+}$ and $D^{0}$, respectively and can be measured\cite{epj:c40:349, Aktas:2004ka} with appropriate silicon trackers such as those at H1  and ZEUS.

Measurements of the $D^{+}$ and $D^{0}$ cross sections are presented with improved precision and in a kinematic region extending to lower transverse momentum, $p_{T}^{D}$, than the previous ZEUS results\cite{Chekanov:2007ch}; this is made possible through the use of the precision tracking provided by the ZEUS microvertex detector (MVD). Single-differential cross sections have been measured as a function of $Q^{2}$, the Bjorken scaling variable, $x$, $p_{T}^{D}$, and the pseudorapidity, $\eta^{D}$, of the $D$ mesons. The cross sections are compared to the predictions of a next-to-leading-order (NLO) QCD calculation using parameterisations of the parton densities in the proton which were determined from fits to inclusive DIS measurements from ZEUS and fixed-target experiments. The cross-section measurements are used to extract the open-charm contribution, $F_{2}^{c\bar{c}}$, to the proton structure function, $F_{2}$.

\section{Experimental set-up}
\label{sec-exp}

The analysis was performed with data taken from 2004 to 2005 when HERA collided 
electrons with energy $E_{e} = 27.5\,\rm GeV$ with protons of energy 
$E_{p} = 920\,\rm GeV$. The results are based on an $e^{-}p$ sample corresponding 
to an integrated luminosity of \mbox{133.6 $\pm$ 3.5\,pb$^{-1}$}.  

A detailed description of the ZEUS detector can be found elsewhere \cite{zeus:1993:bluebook}. A brief outline of the components that are most relevant for this analysis is given below. 

In the kinematic range of the analysis, charged particles were tracked in the central tracking detector (CTD) \cite{nim:a279:290,*npps:b32:181,*nim:a338:254} and the MVD \cite{Brock:2007sw}. These components operated in a magnetic field of 1.43 T provided by a thin superconducting solenoid. The CTD consisted of 72 cylindrical drift chamber layers, organised in nine superlayers covering the polar-angle\footnote{The ZEUS coordinate system is a right-handed Cartesian system, with the $Z$ axis pointing in the proton beam direction, referred to as the ``forward direction'', and the $X$ axis pointing left towards the centre of HERA. The coordinate origin is at the nominal interaction point.} region $15^{\circ} < \theta < 164^{\circ}$. 

The MVD consisted of a barrel (BMVD) and a forward (FMVD) section with three cylindrical layers and four planar layers of single-sided silicon strip sensors in the BMVD and FMVD respectively. The BMVD provided polar-angle coverage for tracks with three measurements from $30^{\circ}$ to $150^{\circ}$. The FMVD extended the polar-angle coverage in the forward region to $7^{\circ}$. After alignment, the single-hit resolution  of the BMVD  was $25\,\mu$m and the impact-parameter resolution of the CTD-BMVD system for high-momentum tracks was $100\,\mu$m.

The high-resolution uranium--scintillator calorimeter (CAL) \cite{nim:a309:77,*nim:a309:101,*nim:a321:356,*nim:a336:23} consisted of three parts: the forward (FCAL), the barrel (BCAL) and the rear (RCAL) calorimeters. Each part was subdivided transversely into towers and longitudinally into one electromagnetic section (EMC) and either one (in RCAL) or two (BCAL and FCAL) hadronic sections (HAC). The smallest subdivision of the calorimeter was called a cell. The CAL energy resolutions, as measured under test-beam conditions, were $\sigma(E)/E = 0.18/\sqrt{E}$ for electrons and $\sigma(E)/E = 0.35/\sqrt{E}$ for hadrons, with $E$ in GeV.

The position of the scattered electron was determined by combining information from the CAL and, where available, the small-angle rear tracking detector (SRTD) \cite{nim:a401:63} and the hadron-electron separator (HES) \cite{nim:a277:176}. 

The luminosity was measured using the Bethe-Heitler reaction $ep \rightarrow e \gamma p$ with the luminosity detector which consisted of independent lead--scintillator calorimeter \cite{desy-92-066,*zfp:c63:391,*acpp:b32:2025} and magnetic spectrometer\cite{physics-0512153} systems. The fractional systematic uncertainty on the measured luminosity was 2.6\%.


\section{Event selection and reconstruction}
\label{SEC:REC}

A three-level trigger system was used to select events 
online~\cite{zeus:1993:bluebook, uproc:chep:1992:222, nim:a580:1257}. At the third level, 
events with a reconstructed scattered-electron or $D$-meson candidate were kept for further analysis.

The kinematic variables $Q^{2}$, $x$ and the fraction of 
the electron energy transferred to the proton in its rest frame, $y$, were reconstructed 
using the double angle (DA) method \cite{proc:hera:1991:23} which relies on the angles 
of the scattered electron and the hadronic energy flow.

The events were selected offline with the following cuts:

\begin {itemize}

\item{$E_{e^{\prime}} >$ 10\,GeV, where $E_{e^{\prime}}$ is the energy of the scattered 
electron;}
\item{$y_{e} < 0.95$, where $y_e$ is determined from the energy and angle of the scattered 
electron. This condition removes events where fake electrons are found in the FCAL;}

\item{ $y_{\rm JB} > 0.02$, where JB signifies the Jacquet-Blondel \cite{proc:epfacility:1979:391} method of kinematic reconstruction. This condition rejects events where the hadronic system cannot be measured precisely;}

\item{$40 <\delta < 65 \, \rm GeV$, where $\delta = \sum E_{i}(1 - \rm cos (\theta_{\it{i}}))$ and $E_{i}$ is the energy of the $i^{\rm th}$ energy-flow object (EFO) \cite{thesis:briskin:1998} reconstructed from tracks detected in the CTD and MVD and energy clusters measured in the CAL. The sum $i$ runs over all  EFOs;}

\item{ $|Z_{\rm vtx}| < 50 \, \rm cm$, where $Z_{\rm vtx}$ is the primary vertex position determined from tracks;}

\item{the impact point ($X, Y$), of the scattered electron on the surface of the RCAL must lie outside the region ($\pm$15 cm, $\pm$15 cm) centred on (0,0).}

\end{itemize} 

Electron candidates in the transition regions between FCAL and BCAL as well as between 
BCAL and RCAL were rejected because of the poor energy reconstruction in these areas.
The angle of the scattered electron was determined using either its impact position on the CAL inner face or a reconstructed track. When available, SRTD and HES were also used. The energy of the scattered electron was corrected for non-uniformity due to geometric effects caused by cell and module boundaries. 

The selected kinematic region was \mbox{$5 < Q^{2} < 1000 \, \rm GeV^{2}$} and \mbox{$0.02 < y < 0.7$}. The production of $D^{+}$  and $D^{0}$ mesons was measured in the range of transverse momentum $1.5 < p_{T}^{D} < 15 \, \rm GeV$ and pseudorapidity $|\eta^{D} | < 1.6$.

The decay-length significance is a powerful variable for the rejection of combinatorial background and is defined as $S_{l} = l/\sigma_{l}$, where $l$ is the decay length in the transverse plane and $\sigma_{l}$ is the uncertainty associated with this distance. The decay length is the distance in the transverse plane between the point of creation and decay vertex of the meson and is given by

\begin{equation}
l = \frac{\left (\vec S_{XY} - \vec{B}_{XY} \right) \cdot \vec p^{D}_{T}}{p_{T}^{D}},
\end{equation}

where $\vec p^{D}_{T}$ is the transverse momentum vector and $\vec {S}_{XY}$ is the two dimensional position vector of the reconstructed decay vertex projected onto the $XY$ plane. The vector $\vec{B}_{XY}$ points to the fitted geometrical centre of the beam-spot which is taken as the origin of the $D$ meson. The centre of the elliptical beam-spot was determined every 2000 well measured events \cite{thesis:nicholass:2008} by fitting a 
Gaussian curve to the $X, Y$ and $Z$ distributions of the primary vertex. 
The mean of these fitted curves was then taken to be the beam-spot position. The 
widths of the beam-spot were 80 $\mu$m and 20 $\mu$m in the 
$X$ and $Y$ directions, respectively.  The decay-length error, $\sigma_{l}$, was 
determined by folding the width of the beam-spot with the covariance matrix of the 
decay vertex after both were projected onto the $D$ meson momentum vector.

\boldmath
\subsection{$D$-meson reconstruction}
\unboldmath

The $D^{+}$ (and $D^{-}$) mesons were reconstructed in the decay channel $D^{+} \rightarrow K^{-} \pi^{+} \pi^{+} (+c.c.)$. In each event, all track pairs with equal charges were combined with a third track with opposite charge to form a $D^{+}$ candidate. The pion mass was assigned to the tracks with equal charges and the kaon mass was assigned to the remaining track. These were then associated and refitted to a common decay vertex \cite{nim:a241:115} and the invariant mass, $M(K\pi\pi)$, was calculated. The tracks were required to have transverse momentum $p_{T}^{\pi} > 0.25$\,GeV and $p_{T}^{K} > 0.5$\,GeV for the pion and kaon tracks, respectively. To ensure that all tracks used were well reconstructed they were required to have passed through 3 superlayers of the CTD and have at least 2 BMVD measurements in the $XY$ plane and 2 in the $Z$ direction.

\begin{figure}
\begin{center}
\includegraphics[width=\textwidth]{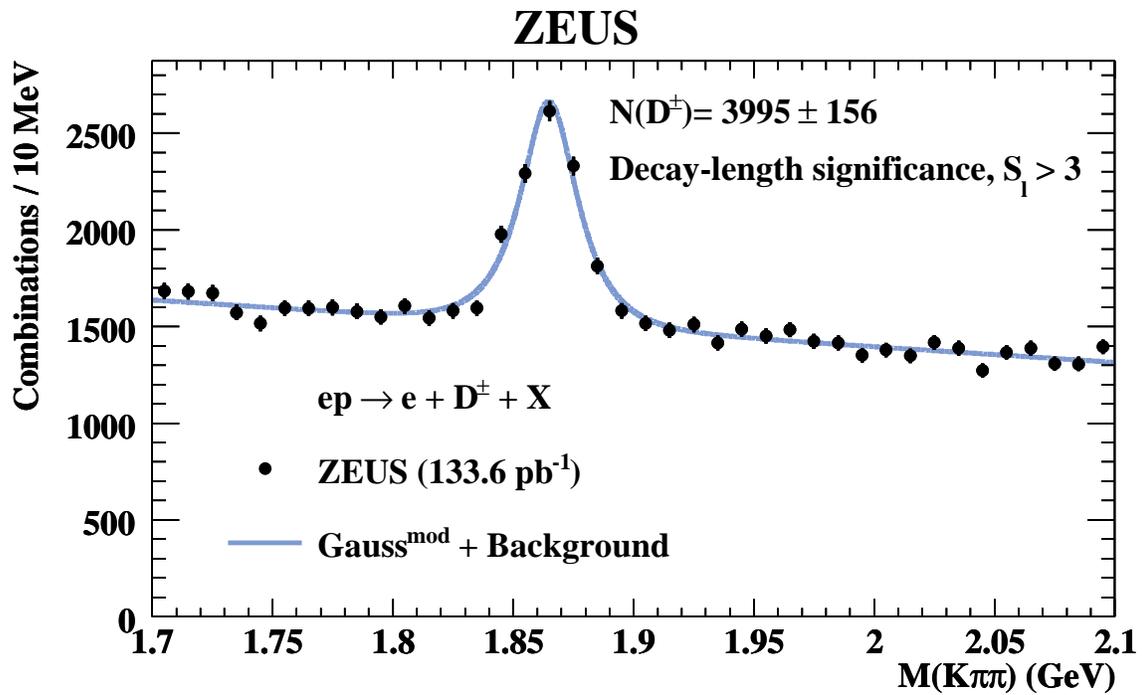}
\caption{The $M(K\pi\pi)$ distribution for the $D^{\pm}$ candidates (dots). The solid curve represents a fit to the sum of a modified Gaussian function and a linear background function.}
\label{FIG:dchsignal}
\end{center}
\end{figure}

\begin{figure}
\begin{center}
\includegraphics[width=0.4\textwidth]{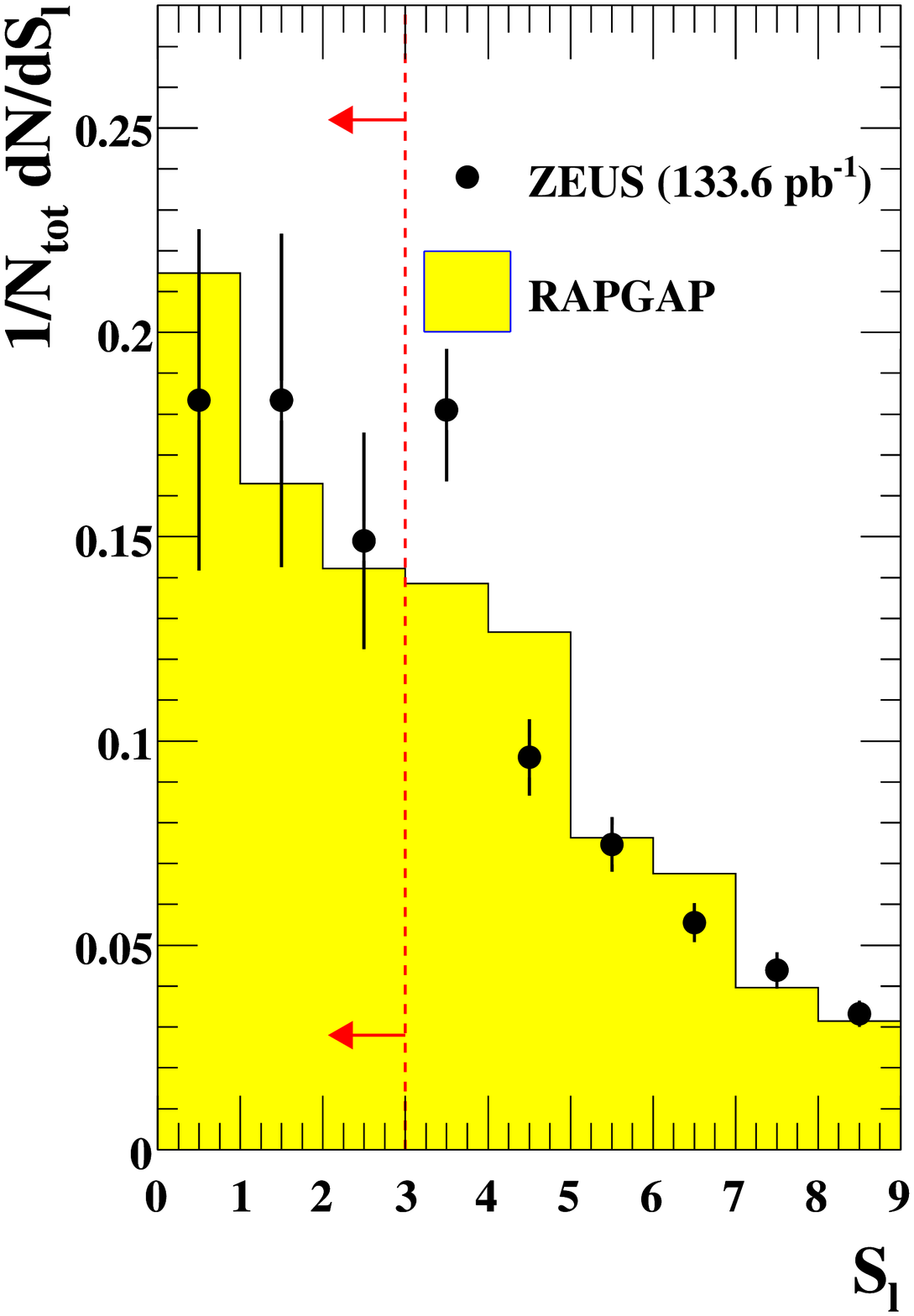}
\put(-35,240){\makebox(0,0)[tl]{\large (a)}}
\includegraphics[width=0.4\textwidth]{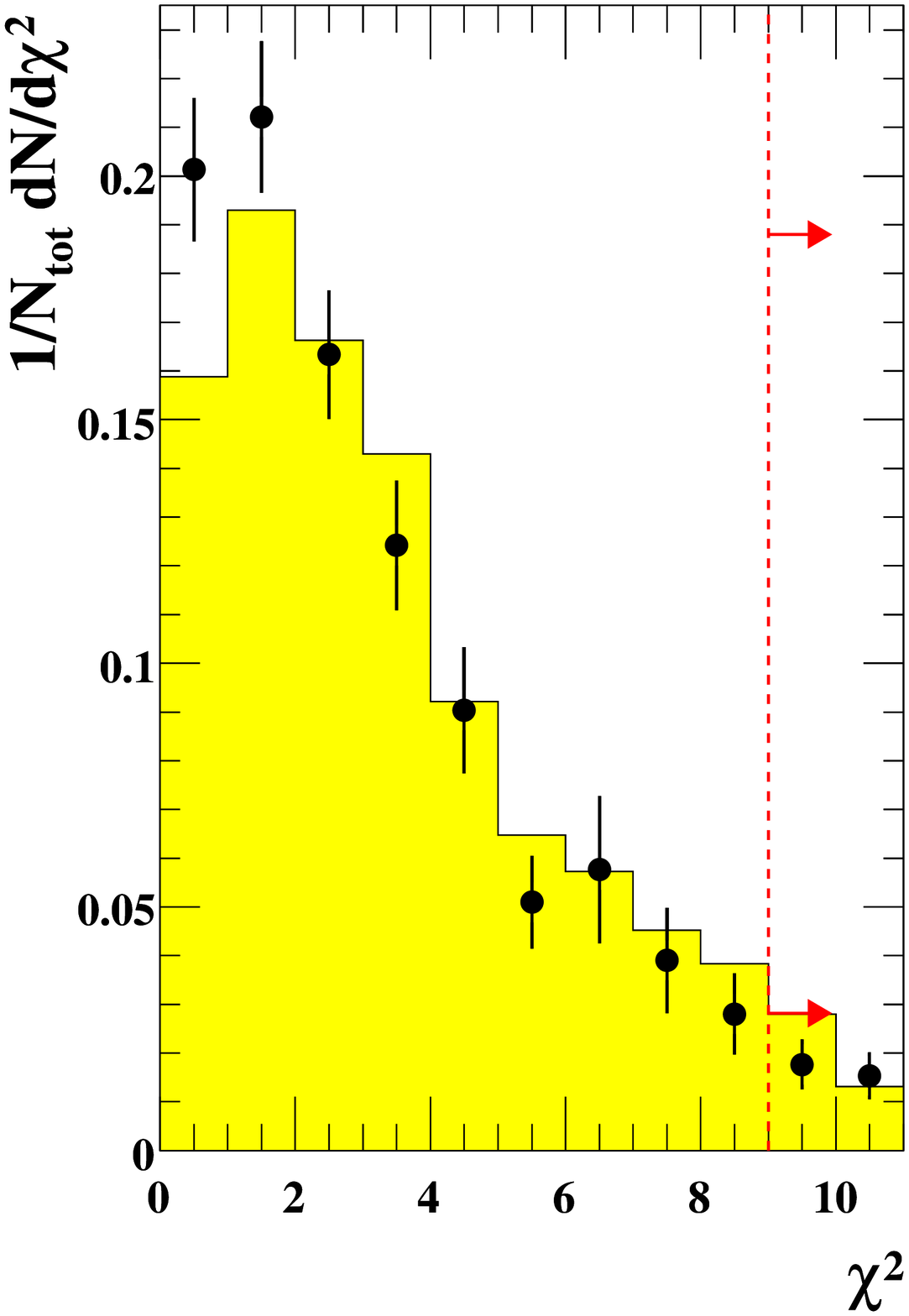}
\put(-34,240){\makebox(0,0)[tl]{\large (b)}}
\put(-205, 280){\makebox(0,0)[tl]{\bf \Huge ZEUS}} \\
\includegraphics[width=0.4\textwidth]{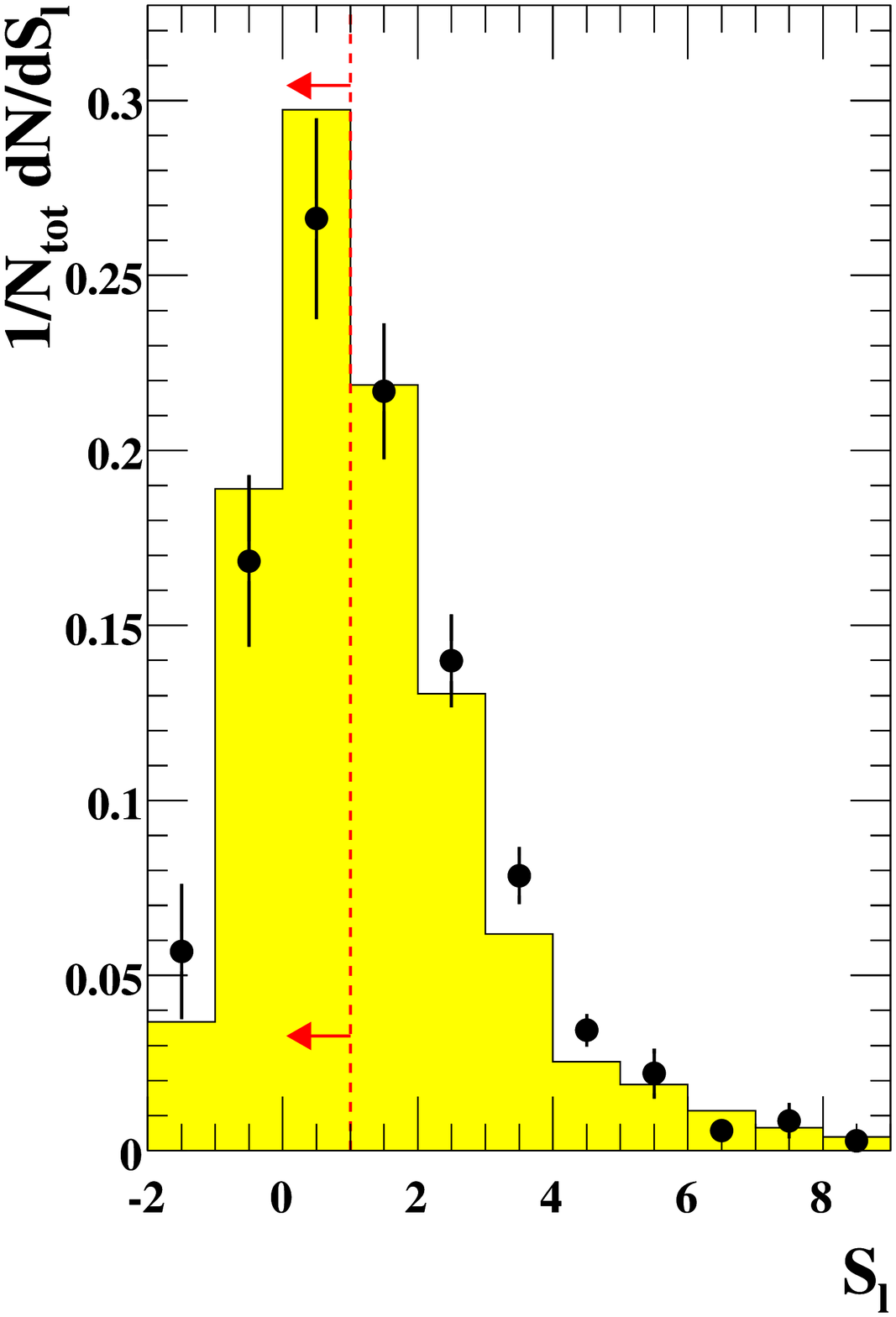}
\put(-35,240){\makebox(0,0)[tl]{\large (c)}}
\includegraphics[width=0.4\textwidth]{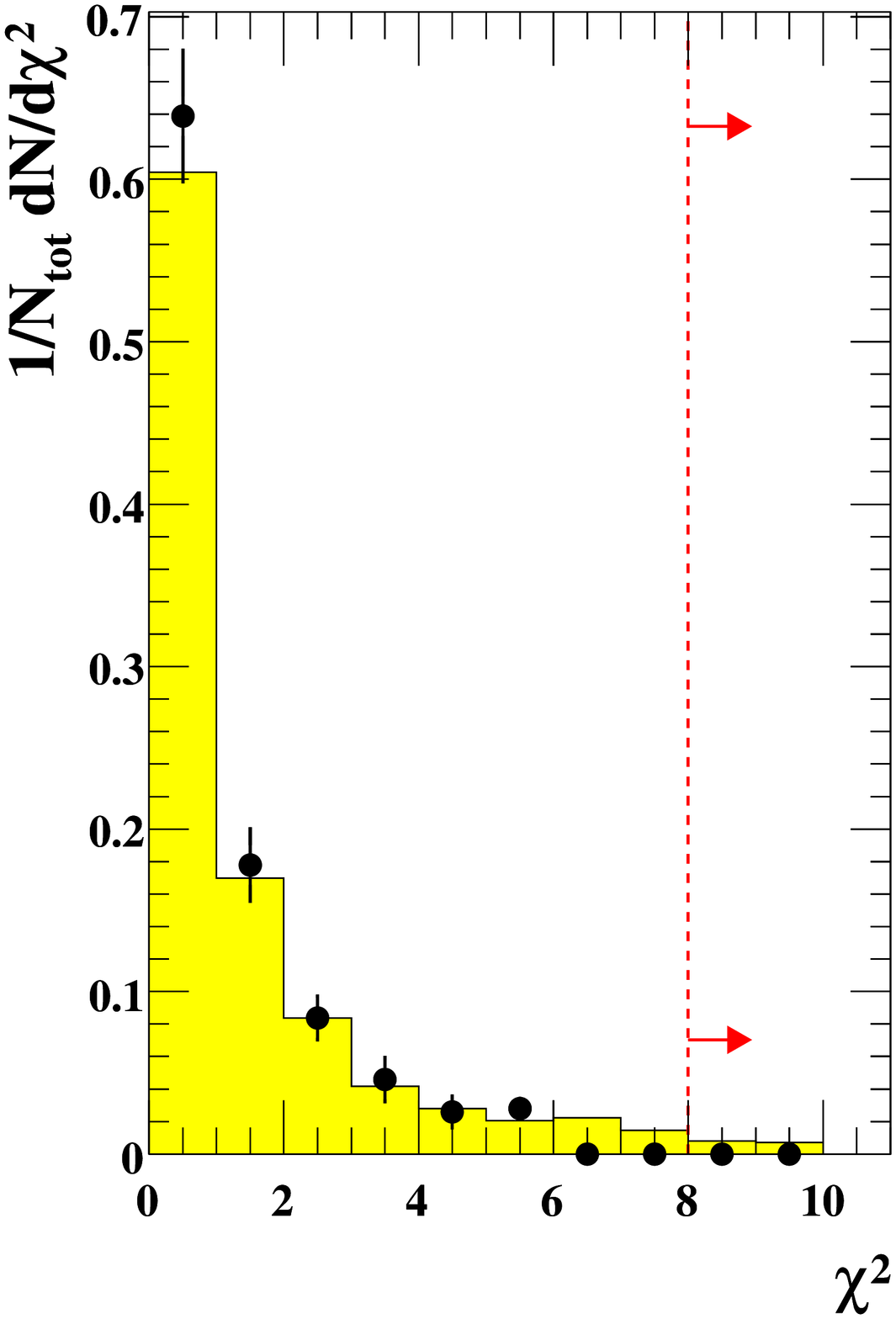}
\put(-35,240){\makebox(0,0)[tl]{\large (d)}}
\caption{Reconstructed decay-vertex variables $S_{l}$ and $\chi^{2}$ for (a, b) $D^{\pm}$ and (c, d) $D^{0}/\bar{D}^{0}$ mesons. Data (points) are compared to detector-level {\sc Rapgap} predictions (shaded histograms). All histograms are normalised to unit area. The dashed line indicates regions removed by the cuts placed on these variables.}
\label{fig:decaycontrol}
\end{center}
\end{figure}

Figure \ref{FIG:dchsignal} shows the $M(K\pi\pi)$ distribution for $D^{+}$ candidates. The combinatorial background was reduced by the requirements that the $\chi^{2}$ of the decay vertex be less than 9 for 3 degrees of freedom and that the decay-length significance, $S_{l}$, be greater than 3 (see Fig. \ref{fig:decaycontrol}). In order to extract the number of reconstructed $D^{+}$ mesons the $M(K\pi\pi)$ distribution was fitted with the sum of a modified Gaussian function \cite{epj:c44:13} and a linear background function. The modified Gaussian function used was

\begin{equation}
{\rm Gauss^{\rm mod}} \propto {\rm exp}  \left [ -0.5 \cdotp x^{1 + 1/(1 + 0.5 \cdotp x)} \right ],
\end{equation}

where $x = |[M(K\pi\pi) - M_{0}]/\sigma|$. This functional form described both the data and MC well. The signal position, $M_{0}$, and the width, $\sigma$, as well as the numbers of $D^{+}$ mesons in each signal were free parameters of the fit. The number of reconstructed $D^{+}$ mesons yielded by the fit was $N(D^{+}) = 3995 \pm 156$.

A sample of $D^{+}$ candidates with $p_{T}^{\pi} > 0.5$\,GeV, $p_{T}^{K} > 0.7$ GeV and 
$p_{T}^{D^{+}} >$ 3\,GeV was used to obtain the lifetime of the $D^{+}$ meson. The 
higher $p_{T}$ cuts were used to obtain a signal with no requirements made 
on the significance of the decay length. The number of reconstructed $D^{+}$ mesons 
yielded by the fit to the data was \mbox{$N(D^{+}) = 4383 \pm 353$.}

The $D^{0}$ (and $\bar{D}^{0}$) mesons were reconstructed in the decay channel $D^{0} \rightarrow K^{-}\pi^{+} (+c.c.)$, with candidates found in a similar manner to the $D^{+}$, except that only oppositely charged pairs of tracks were combined together to form the meson candidate. The tracks were required to have transverse momentum $p_{T}^{K} > 0.7$\,GeV and $p_{T}^{\pi} > 0.3$\,GeV for the kaon and pion tracks. The $\chi^{2}$ and $S_{l}$ cuts were 8 and 1, respectively, with 1 degree of freedom in the vertex fit (see Fig. \ref{fig:decaycontrol}). After selection, the $D^{0}$ candidates were separated into tagged and antitagged samples with the antitagged sample used for cross-section measurements.

\begin{figure}
\begin{center}
\includegraphics[width=\textwidth]{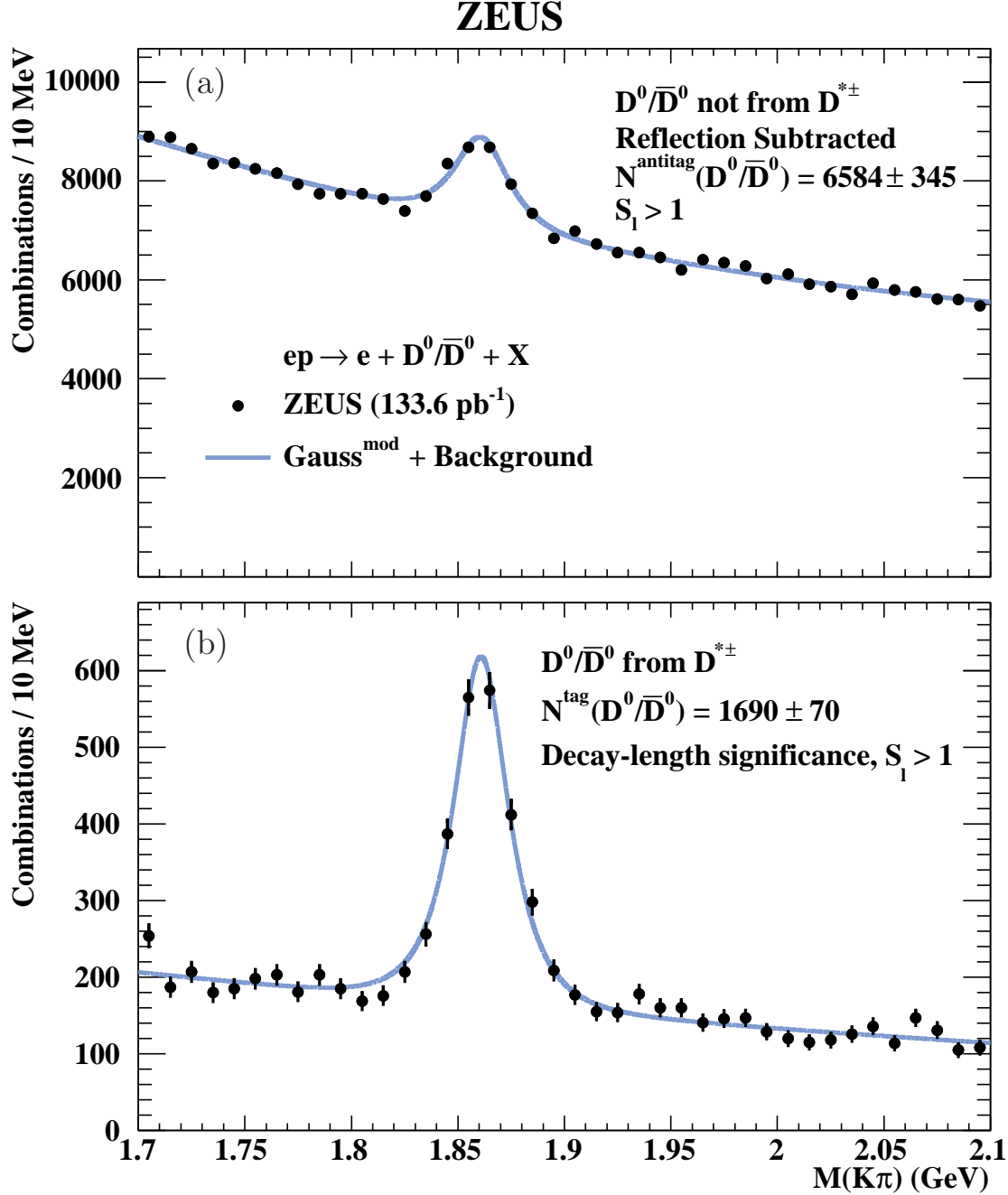}
\put(-365,490){\makebox(0,0)[tl]{\large (a)}}
\put(-365,255){\makebox(0,0)[tl]{\large (b)}}
\caption{The $M(K\pi)$ distributions (dots) for (a) $D^{0}/\bar{D}^{0}$ candidates not consistent with a $D^{*\pm}$ decay, obtained after the reflection subtraction (see text) and (b) $D^{0}/\bar{D}^{0}$ candidates consistent with a $D^{*\pm}$ decay. The solid curves represent the simultaneous fit as described in the text.} 
\label{FIG:d0signal}
\end{center}
\end{figure}

The tagged group consisted of $D^{0}$ candidates which are consistent with a $D^{*\pm} \rightarrow D^{0}\pi_{s}^{\pm}$ decay when combined with a third track that could be a ``soft'' pion, ($\pi_{s}$). The soft pion was required to have $p_{T} > 0.12$\,GeV and charge opposite to that of the kaon. The tagged $D^{0}$ sample was used for the correction of the MC and reflection subtraction in the antitagged sample. For the antitagged sample, containing $D^{0}$ mesons not coming from a $D^{*\pm}$, incorrect assignment of the pion and kaon masses produced a wider reflected signal. The distribution of this reflection was estimated using the tagged $D^{0}$ candidates and, after normalising it to the ratio of the number of $D^{0}$ mesons in the two samples, it was subtracted from the antitagged $D^{0}$ candidates. Figure \ref{FIG:d0signal} shows the $M(K\pi)$ distributions for tagged and antitagged $D^{0}$ candidates. The distributions were fitted simultaneously assuming that both have the same peak position and width and, like the $D^{+}$, were parameterised as a modified Gaussian function. The number of antitagged (tagged) $D^{0}$ mesons yielded by the fit was  $N^{\rm antitag}(D^{0}) = 6584 \pm 345 \, (N^{\rm tag}(D^{0}) = 1690 \pm 70)$.

A sample of $D^{0}$ candidates with $p_{T}^{\pi, K} > 0.8$\,GeV and $p_{T}^{D^{0}} > 3$\,GeV was used to obtain the lifetime of the $D^{0}$ meson. The higher $p_{T}$ cuts were used to obtain a signal with no requirements made on the significance of the decay length. The number of antitagged (tagged) $D^{0}$ mesons yielded by the fit was $N^{\rm antitag}(D^{0}) = 5612 \pm 283$ \mbox{$(N^{\rm tag}(D^{0}) = 1495 \pm 56)$.}

\boldmath
\subsection{$D$-meson lifetimes}
\unboldmath

Lifetimes for the $D^{+}$ and $D^{0}$ mesons were calculated using decay lengths in the transverse plane and reconstructed $D$-meson signals in the kinematic region  $5 < Q^{2} < 1000 \, \rm GeV^{2}$, 
$0.02 < y < 0.7$, $3 < p_{T}^{D} < 15\, \rm GeV$ and $|\eta^{D}| < 1.6$. Unfolding is not necessary as the detector acceptance is uniform with respect to 
the displacement of the secondary vertex and the normalisation of the lifetime 
distribution is irrelevant. The number of $D$ mesons in a given bin of proper decay length, 
$ct$, was extracted and the distributions fitted with the function

\begin{equation}
f(ct) = \frac{1}{2\lambda}{\rm exp}\left[{-\left(\frac{ct}{\lambda}-\frac{\sigma^{2}}{2\lambda^{2}} \right)}\right]\int_{u_{\rm min}}^{\infty} e^{-u^{2}} \, du, 
\end{equation} 
where $u_{\rm min} =\left (-ct/\sigma + \sigma/\lambda \right)$, $\lambda$ is the lifetime and $\sigma$ is the spatial resolution. This function represents an exponential decay convoluted with a Gaussian resolution. For the purposes of the lifetime extraction, $\sigma$ was set to the value extracted from the tagged $D^{0}$ sample, 120 $\mu$m, which depended only weakly on $p_{T}$.

\begin{figure}
\begin{center}
\includegraphics[width=\textwidth]{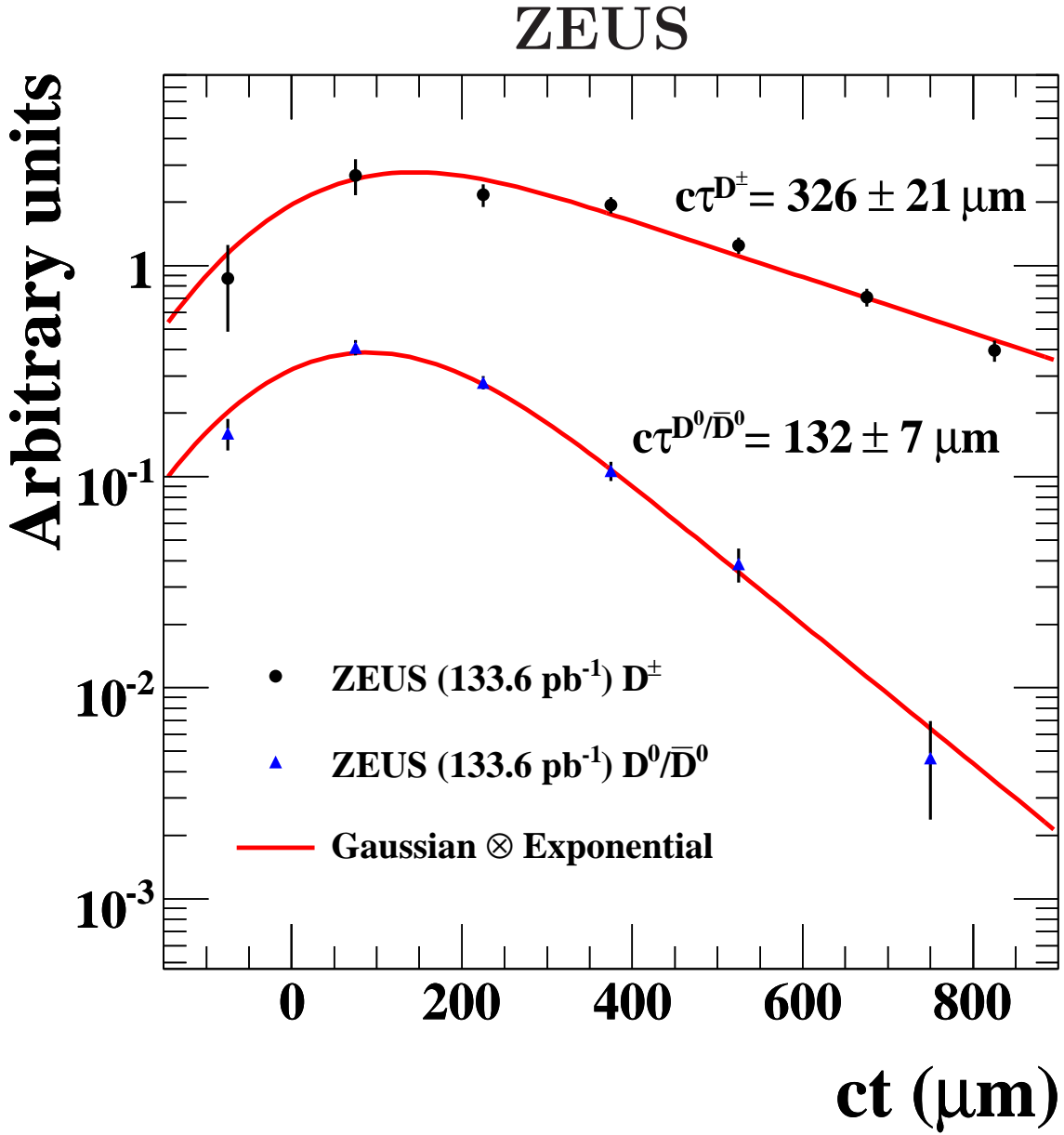}
\put(-245, 460){\makebox(0,0)[tl]{\bf \Huge ZEUS}} 
\caption{The distributions of reconstructed $D^{\pm}$ candidates (circles) and $D^{0}/\bar{D}^{0}$ candidates not consistent with a $D^{*\pm}$ decay (triangles) extracted in bins of proper decay length, $ct$. Both distributions are fitted with functions described by a Gaussian convoluted with an exponential decay. The relative normalisation of the distributions is chosen to aid visibility.}
\label{fig:life}
\end{center}
\end{figure}

The fitted $ct$ distributions for $D^{+}$ and $D^{0}$ mesons are shown in Fig. \ref{fig:life} and the extracted values for the lifetime are:
\begin{eqnarray}
c\tau (D^{+}) &=& 326 \pm 21 (\rm stat.) \, \mu m \nonumber \\
c\tau (D^{0})   &=& 132 \pm  7 (\rm stat.) \, \mu m \nonumber
\end{eqnarray}
The systematic uncertainties are significantly smaller than the statistical 
uncertainty as the measurement has only a small dependence on the details of the MC 
simulation. The values are consistent with the world average values of \mbox{311.8\,$\pm$\,2.1\,$\mu$m} and 
\mbox{122.9\,$\pm$\,0.5\,$\mu$m}~\cite{PDBook} for the $D^{+}$ and $D^{0}$, respectively.


\section{Monte Carlo models}
\label{sec:mc}

The acceptances were calculated using the {\sc Rapgap} 3.00 \cite{cpc:86:147} Monte Carlo (MC) model, which was interfaced with {\sc Heracles} 4.6.1 \cite{cpc:69:155} in order to 
incorporate first-order electroweak corrections. The generated events were then passed through a full simulation of the detector using {\sc Geant} 3.21~\cite{tech:cern-dd-ee-84-1} before being processed and selected with the same software as used for the data.
 
The MC was used to simulate events containing charm produced by the BGF process. The {\sc Rapgap} generator used leading-order matrix elements with leading-logarithmic parton-shower radiation. The CTEQ5L \cite{epj:c12:375} PDF for the proton was used, and the charm quark mass was set to 1.5\,GeV. Charm fragmentation was implemented using the Lund string model \cite{prep:97:31}. $D$ mesons originating from $B$ decays were accounted for by inclusion of a {\sc Rapgap} $b$-quark sample where the $b$-quark mass was set to 4.75 GeV.

A weighting procedure utilising the tagged $D^{0}$ sample was applied in order to correct for imperfections in the MC description of the decay-length uncertainty \cite{thesis:nicholass:2008}.
 

\section{NLO QCD calculations}
\label{SEC:THEORY}

The NLO QCD predictions for the $c\bar{c}$ cross sections were obtained using the HVQDIS program \cite{pr:d57:2806} based on the fixed-flavour-number scheme (FFNS). In this scheme, only light partons ($u, d, s,$ and $g$) are included in the initial-state proton as partons whose $x, Q^{2}$ distributions obey the DGLAP equations \cite{sovjnp:15:438,*sovjnp:20:94,*np:b126:298,*jetp:46:641} and the $c\bar{c}$ pair is produced via the BGF mechanism with NLO corrections \cite{np:b452:109,*pl:b353:535,*np:b392:162,*np:b392:229}. The presence of different large scales, $Q$, $p_{T}$ and the mass of the $c$ quark, $m_{c}$, can spoil the convergence of the perturbative series because the neglected terms of orders higher than $\alpha_{s}^{2}$ (where $\alpha_{s}$ is the strong coupling constant) contain log$(Q^{2}/m_{c}^{2})$ factors which can become large. 

The predictions for $D$-meson production at NLO were obtained using HVQDIS with the following inputs. The ZEUS-S NLO QCD global fit \cite{pr:d67:012007} to 
structure function data was used as the parameterisation of the proton PDFs. This fit was repeated~\cite{misc:www:zeus2002} in the FFNS, in which the PDF has three active quark 
flavours in the proton. In this fit $\Lambda^{(3)}_{\rm QCD}$ was set to 0.363\,GeV and the mass of the charm quark was set to 1.5\,GeV; the same mass was therefore used in the 
HVQDIS calculation. The renormalisation and factorisation scale, $\mu = \mu_{R} = \mu_{F}$, was set to $\sqrt{Q^{2} + 4m_{c}^{2}}$. The charm fragmentation to the particular $D$ meson was described by the Peterson function \cite{pr:d27:105} with the Peterson parameter, $\epsilon$, set to 0.035~\cite{Nason:1999zj}. The values used for the hadronisation 
fractions, $f(c \rightarrow D)$, were those previously measured in DIS at ZEUS, 0.216$^{+0.021}_{-0.029}$ and 0.450$^{+0.027}_{-0.060}$ for the $D^{+}$ and antitagged $D^{0}$, respectively~\cite{Chekanov:2007ch}. 

To estimate the contribution of beauty production, the HVQDIS calculation and hadronisation from the MC were combined, using $d\sigma(b\rightarrow D)_{\rm NLO+MC} = d\sigma(b\bar{b})_{\rm NLO} \cdotp \mathcal{C}_{\rm had}$ where $\mathcal{C}_{\rm had} = d\sigma(b\rightarrow D)_{\rm MC}/d\sigma(b\bar{b})_{\rm MC}$. The ZEUS NLO QCD fit was used as the proton PDF, so that the mass used in this fit was also used in the HVQDIS program. The hadronisation fraction, $f(b\rightarrow D)$, was set to 0.231 and 0.596 for the $D^{+}$ and $D^{0}$, respectively \cite{Buskulic:1996ah}.   

The HVQDIS predictions for $D$-meson production are affected by theoretical uncertainties listed below. The average uncertainty on the total cross sections is given in parentheses: 

\begin{itemize}

\item{the ZEUS PDF uncertainties propagated from the experimental uncertainties of the fitted data ($\pm 5\%$). The change in the cross section was independent of the kinematic region;}

\item{the mass of the charm quark ($\pm 8\%$). The charm quark mass was changed consistently in the PDF fit and in HVQDIS by $\mp 0.15 \, \rm GeV$;}

\item{the renormalisation and factorisation scale, $\mu$ ($^{+7\% }_{-0\%}$). The scales $2\sqrt{Q^{2}+4m_{c}^{2}}$ and $\sqrt{Q^{2}/4+m_{c}^{2}}$  were used;}

\item{the $\epsilon$ parameter of the Peterson fragmentation function ($^{+5\%}_{-7\%}$) was varied by $^{+0.035}_{-0.015}$ \cite{unpub:fragpaper}.}

\end{itemize}


\section{Data correction and systematic uncertainties}
\label{sec:corr_syst}

For a given observable $Y$, the production cross section was determined using:
\begin{equation}
\frac{d\sigma}{dY} = \frac{N(D)}{\mathcal{A} \cdot \mathcal{L} \cdot \mathcal{B} \cdot \Delta Y}
\end{equation}
where $N(D)$ is the number of reconstructed $D$ mesons in a bin of size $\Delta Y$, $\mathcal{A}$ is the reconstruction acceptance as found from the MC sample which includes migrations, efficiencies and QED radiative effects for that bin, $\mathcal{L}$ is the integrated luminosity and $\mathcal{B}$ is the branching ratio for the decay channel used in the reconstruction. 

\begin{figure}
\begin{center}
\includegraphics[width=0.4\textwidth]{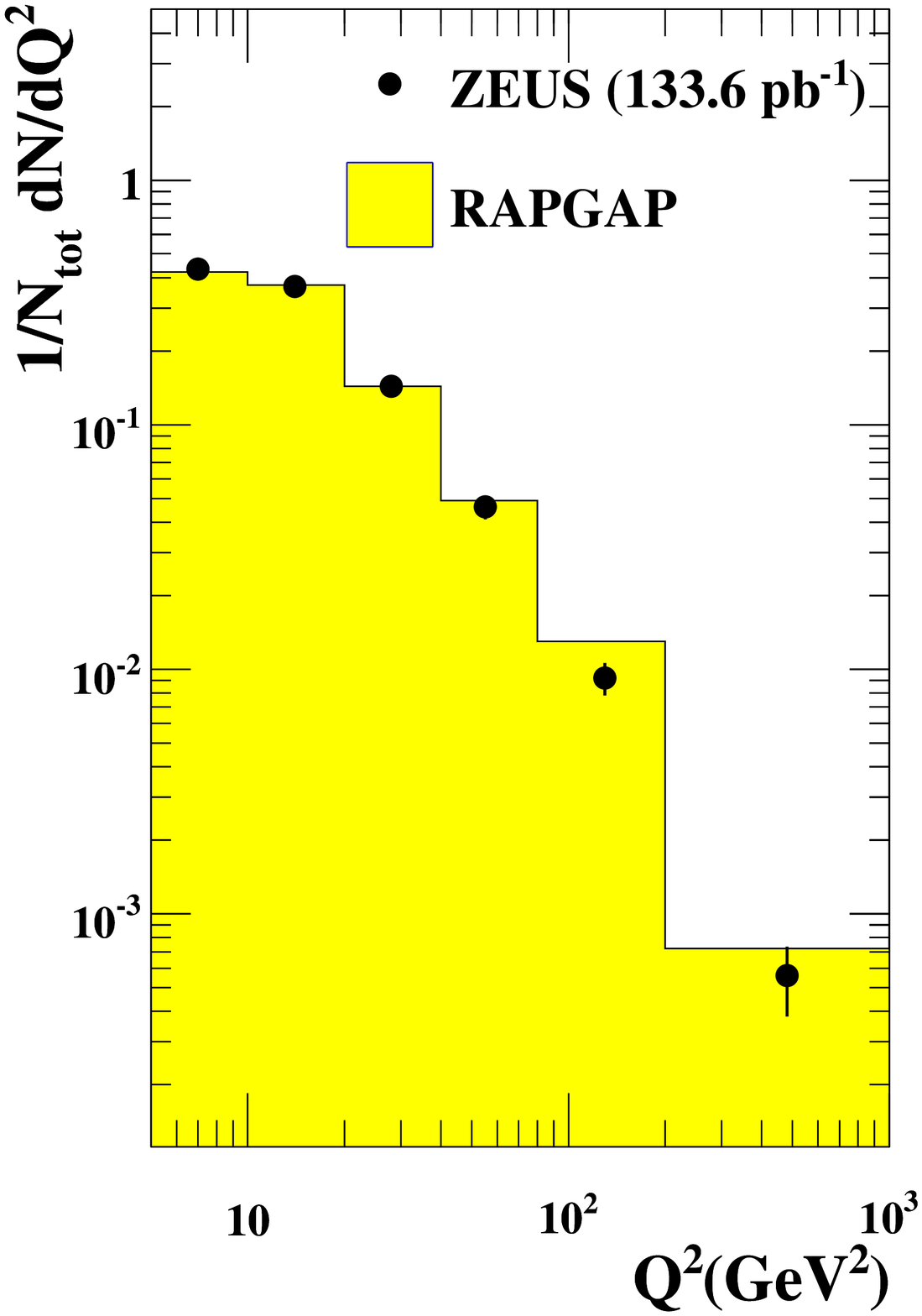}
\put(-140,240){\makebox(0,0)[tl]{\large (a)}}
\includegraphics[width=0.4\textwidth]{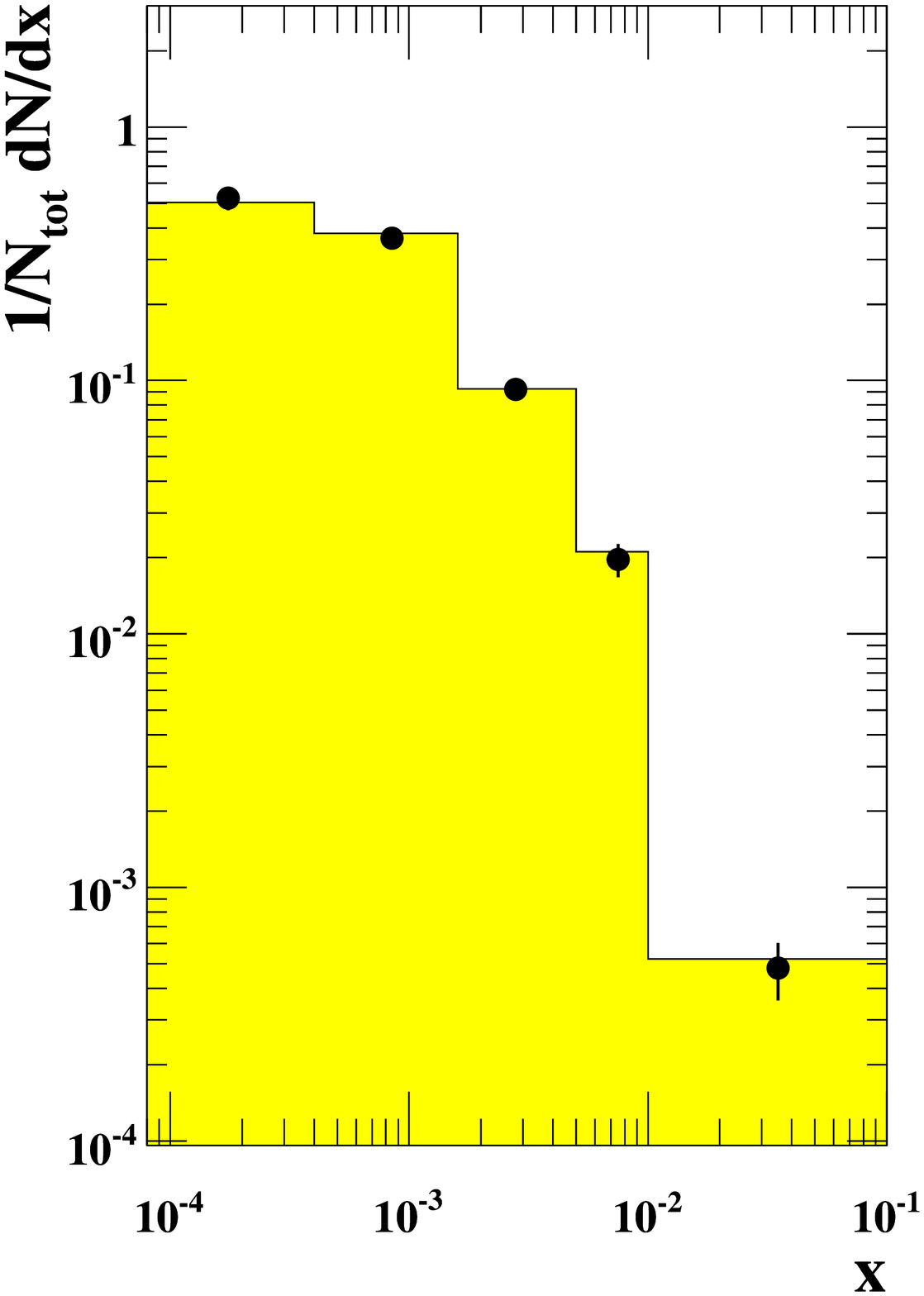}
\put(-140,240){\makebox(0,0)[tl]{\large (b)}}
\put(-205, 280){\makebox(0,0)[tl]{\bf \Huge ZEUS}}  
\caption{Reconstructed DIS variables for events with $D^{\pm}$ candidates (extracted with a fitted signal) for data (points) compared to detector-level {\sc Rapgap} predictions (shaded histograms). All histograms are normalised to unit area. Similar agreement was observed for the $D^{0}/\bar{D}^{0}$ candidates.}
\label{fig:discontrol}
\end{center}
\end{figure}

\begin{figure}
\begin{center}
\includegraphics[width=0.4\textwidth]{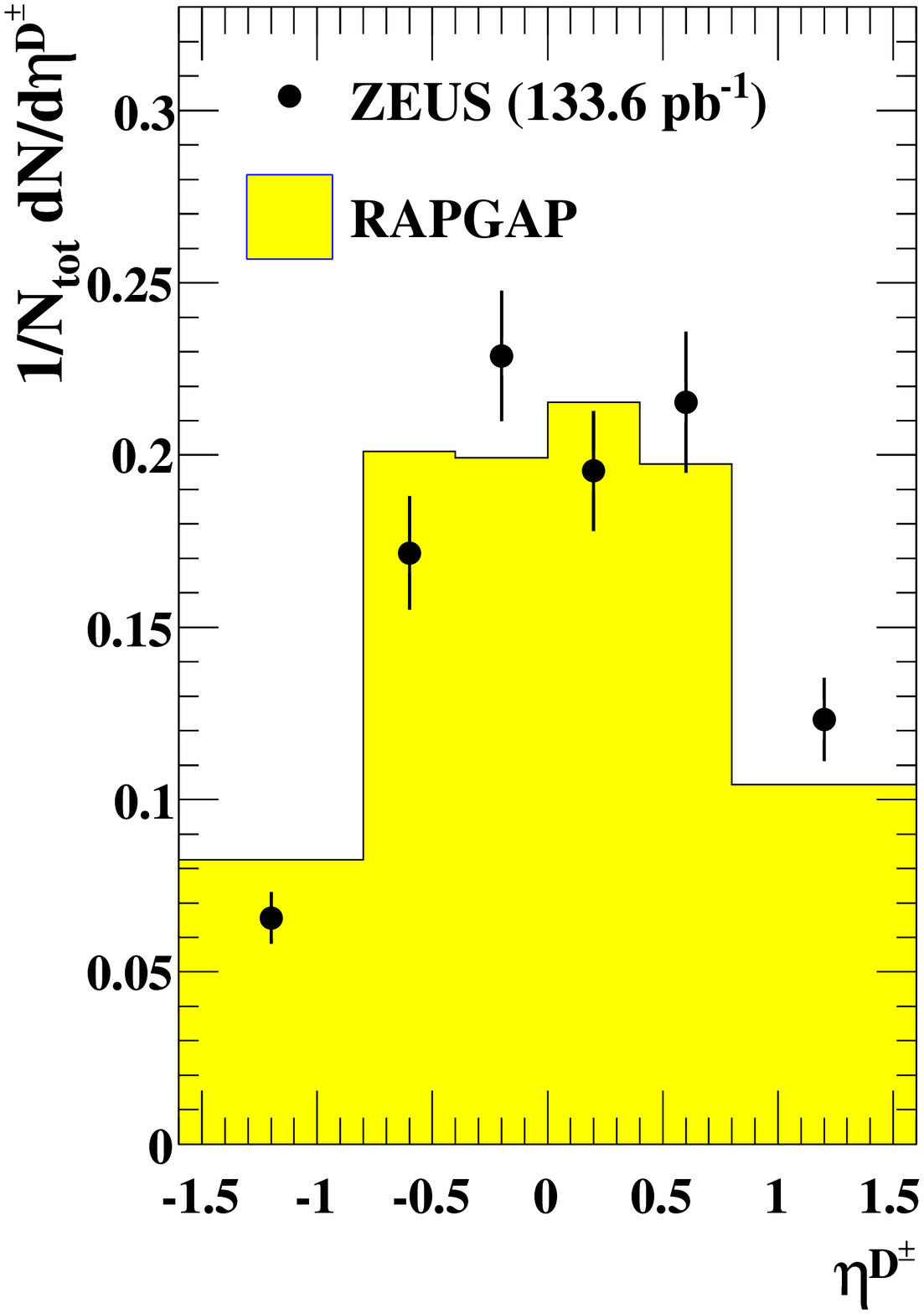}
\put(-35,240){\makebox(0,0)[tl]{\large (a)}}
\includegraphics[width=0.4\textwidth]{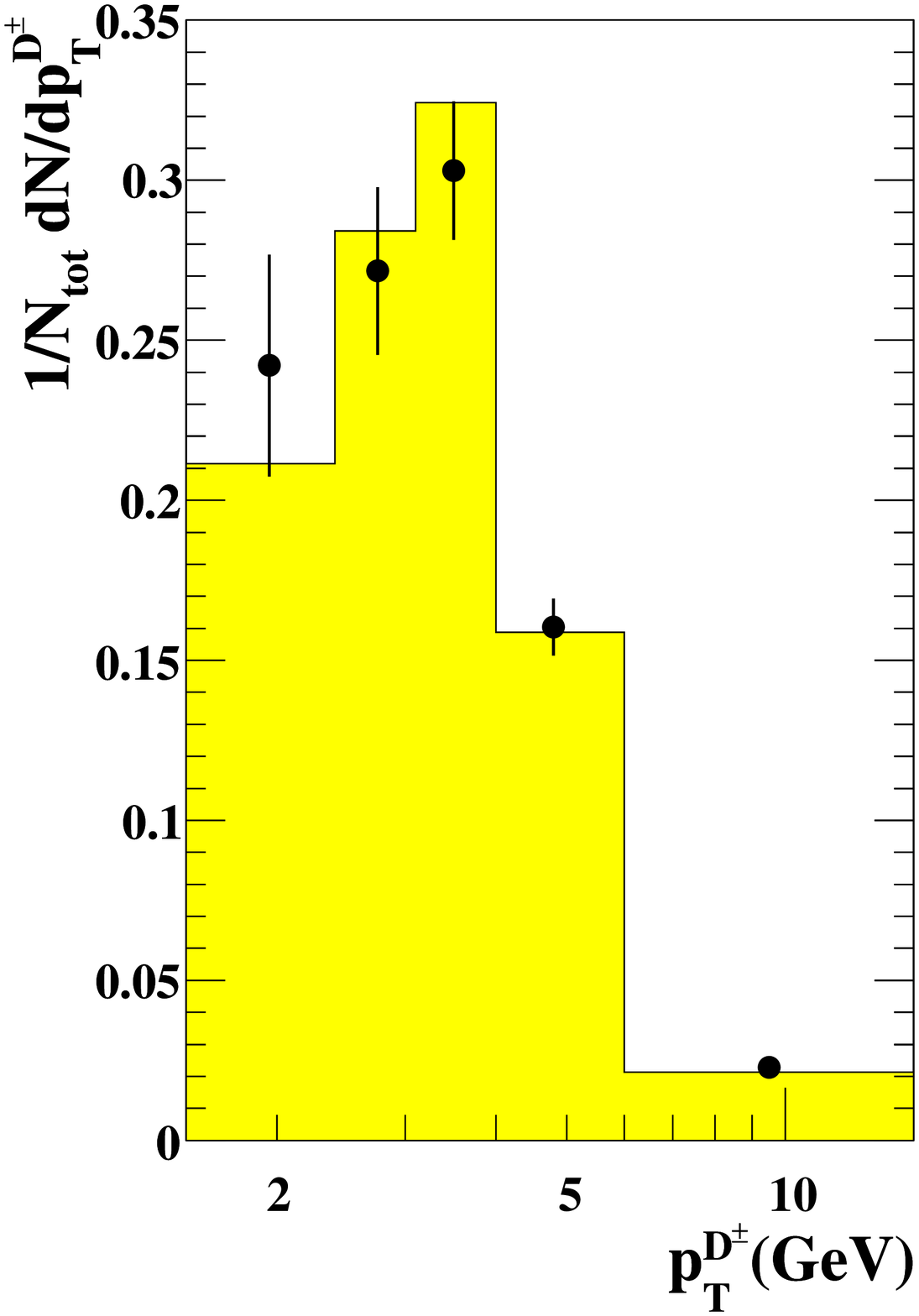}
\put(-35,240){\makebox(0,0)[tl]{\large (b)}} 
\put(-205, 280){\makebox(0,0)[tl]{\bf \Huge ZEUS}} \\
\includegraphics[width=0.4\textwidth]{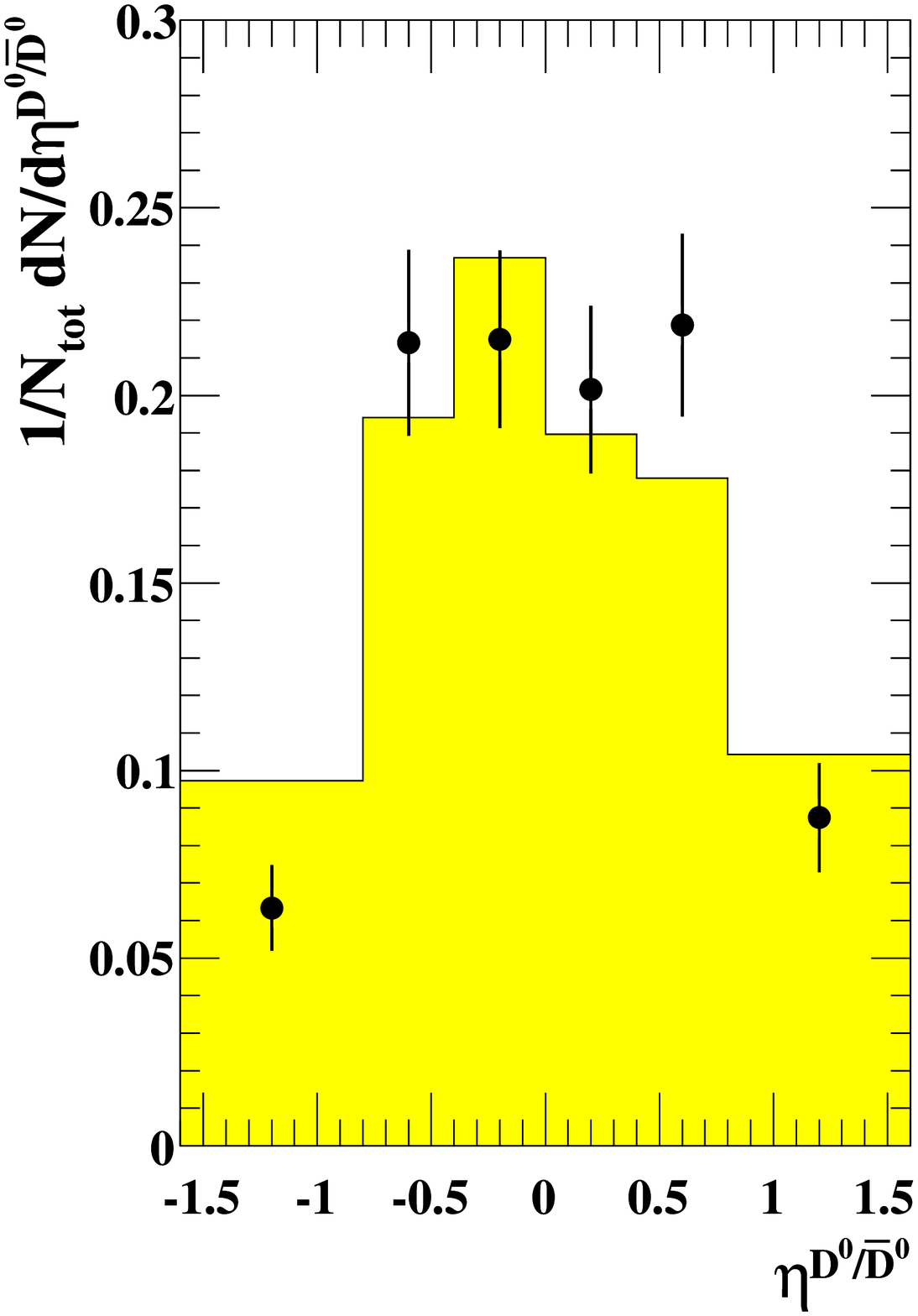}
\put(-35,240){\makebox(0,0)[tl]{\large (c)}}
\includegraphics[width=0.4\textwidth]{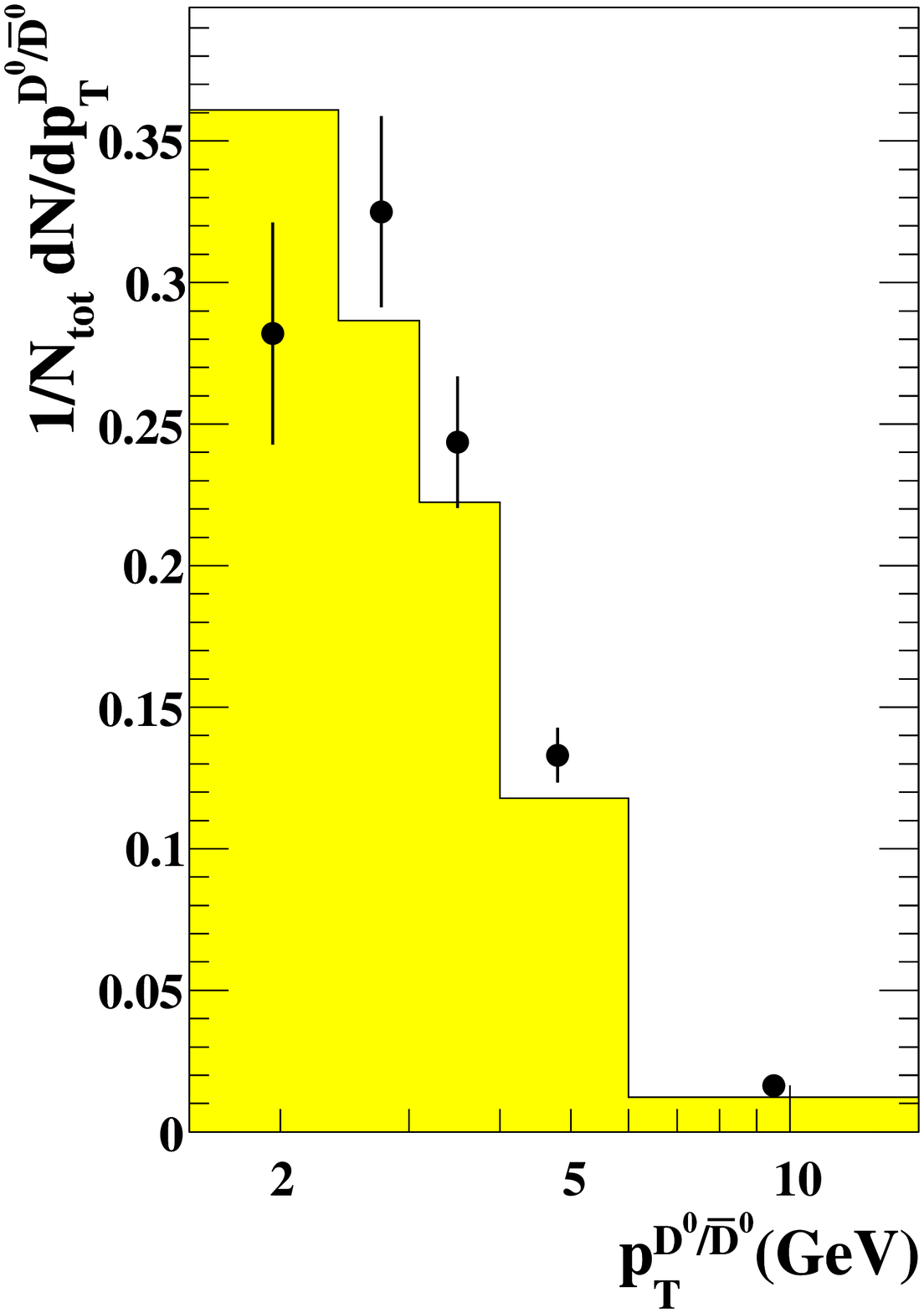}
\put(-35,240){\makebox(0,0)[tl]{\large (d)}}
\caption{Reconstructed (a, b) $D^{\pm}$ and (c, d) $D^{0}/\bar{D}^{0}$ kinematic variables for data (points) compared to detector-level {\sc Rapgap} predictions (shaded histograms). All histograms are normalised to unit area.}
\label{fig:dkinecontrol}
\end{center}
\end{figure}

Small admixtures to the reconstructed signals from other decay modes were taken into account in the MC sample used for the acceptance-correction procedure. To correct from 
the number of reconstructed $D^{0}$ mesons to the production cross sections, small migrations were taken into account between the tagged and antitagged samples. It was checked that the {\sc Rapgap} MC sample gives a reasonable description of the data for selected DIS and $D$ meson variables. Figures \ref{fig:decaycontrol}, \ref{fig:discontrol} and~\ref{fig:dkinecontrol} show important variables for the secondary vertex reconstruction, distributions for the DIS variables and the kinematics of the $D$ meson, respectively. For all variables, the number of reconstructed $D$ mesons is extracted by fitting the number of $D$ mesons in each bin of the distribution. The MC provides a good enough description of the data for acceptance calculations in all variables. 

Reconstruction acceptances vary depending on the particle and kinematic region of the measurement. For example, the overall $D^{+}$ and $D^{0}$ acceptances calculated with 
{\sc Rapgap} after applying the selection criteria for the kinematic region are $\approx 7\%$ and $\approx 17\%$, respectively. The lower average acceptance in relation to previous ZEUS measurements is accounted for by reduced efficiency for reconstructed $D$ mesons due to the extension of the kinematic range to lower $p_{T}^{D}$ and the use of lifetime tagging. This is offset by a gain of a factor of 20 and 3 in the signal to background ratios of the $D^{+}$ and $D^{0}$ samples.

The systematic uncertainties of the measured cross sections were determined by changing the analysis procedure and repeating all calculations. The following possible sources of systematic uncertainties were considered~\cite{thesis:grigorescu:2008,thesis:nicholass:2008} with the average effect on the measured $D^{+}$ and $D^{0}$ total cross sections shown in parentheses:

\begin{itemize}

\item{$\{ \delta_{1} \}$ the cut on $y_{\rm JB}$ was changed by $^{+0.04}_{-0.02}$ ($^{+0.3\%}_{-3\%}$);}

\item{$\{ \delta_{2} \}$ the cut on the scattered electron energy, $E_{e^{\prime}}$,  was changed by $\pm$1\,GeV ($^{+2\%}_{-1\%}$);}

\item{$\{ \delta_{3} \}$ the $b$-quark cross section was varied by a factor 
                         of two in the reference MC sample ($^{+1.3\%}_{-1.7\%}$);}

\item{$\{ \delta_{4} \}$ the uncertainty of the tracking performance was obtained by varying all momenta by $\pm 0.3 \%$ which corresponds to the uncertainty in the magnetic field; and by changing the track momentum and angular resolutions by $^{+20\%}_{-10\%}$ of their values. The asymmetric resolution variations were used since the MC signals typically had somewhat narrower widths than those observed in the data ($\pm 1\%$);}

\item{$\{ \delta_{5} \}$ the uncertainty of the MVD hit efficiency was obtained~\cite{thesis:nicholass:2008} by evaluating the relative difference in single-track efficiency between data and MC when 2 $XY$ and 2 $Z$ measurements were required in the BMVD ($\pm 1.1\%$);}

\item{$\{ \delta_{6} \}$ the cut on $S_{l}$ was varied by $\pm 1.0$ in the $D^{+}$ analysis and $\pm 0.4$ in the $D^{0}$ analysis~\cite{thesis:gassner:2002} ($^{+6\%}_{-7\%}$);}

\item{$\{ \delta_{7} \}$ the cut on the $\chi^{2}$ of the secondary vertex  was changed by $\pm$2 in the $D^{+}$ analysis and $\pm$1.5 in the $D^{0}$ analysis ($^{+2\%}_{-1\%}$);}

\item{$\{ \delta_{8} \}$ the MC $p_{T}^{D}$ distribution was reweighted in order to 
account for the difference (see Fig.~\ref{fig:dkinecontrol}) between data and MC ($<1\%$);}

\item{$\{ \delta_{9} \}$ the MC $\eta^{D}$ distribution was reweighted in order to 
account for the difference (see Fig.~\ref{fig:dkinecontrol}) between data and MC ($<1\%$).}

\end{itemize} 

An additional source of systematic uncertainty in the $D^{0}$ analysis was investigated:

\begin{itemize}

\item{$\{ \delta_{10} \}$ the background function was parameterised by an exponential function ($\pm  4\%$).}

\end{itemize}

Several other sources of systematic uncertainty were considered and found to have an effect of $<1\%$ on the total cross sections. These sources were related to the DIS selection criteria and the method for extracting the number of tagged $D^{0}$ mesons. 

The systematic uncertainty is dominated by $\delta_{6}$, which is related to the description of the MVD resolution. This uncertainty was evaluated from the differences between the data and MC description of $\sigma_{l}$ (see Fig.~\ref{fig:decaycontrol}). This difference was then propagated to a cut variation of the $S_{l}$ cut and the analysis procedure repeated.

Contributions from the different systematic uncertainties were calculated and added in quadrature separately for positive and negative variations. Uncertainties due to those on the luminosity measurement and branching ratios were only included in the measured $D^{+}$ and $D^{0}$ total cross sections. For differential cross sections these uncertainties are not included in the tables and figures.


\section{Cross sections}

Charm meson cross sections for the process $ep \rightarrow eDX$ were calculated using the reconstructed $D^{+}$ and $D^{0}$ signals (see Section \ref{SEC:REC}) in the kinematic region $5 < Q^{2} < 1000 \, \rm GeV^{2}$, $0.02 < y < 0.7$, $1.5 < p_{T}^{D} < 15\, \rm GeV$ and $|\eta^{D}| < 1.6$.

The following cross sections were measured:

\begin{itemize}
\item{The production cross section for $D^{+}$ and $D^{-}$ mesons:
\begin{equation}
\sigma(D^{+}) = 4.67 \pm 0.26  ~(\rm stat.) ~ ^{+0.38}_{-0.56} ~(\rm syst.) \pm 0.17 (\rm br.) \pm 0.12 (\rm lumi.) \rm ~nb \nonumber \\ \
\end{equation}
}
\item{The production cross section for $D^{0}$ and $\bar{D}^{0}$ mesons not originating from $D^{*\pm}$ decays:
\begin{equation}
\sigma^{\rm antitag}(D^{0}) =  7.49 \pm 0.46  ~(\rm stat.) ~ ^{+0.98}_{-0.58} ~(\rm syst.) \pm 0.14 (\rm br.) \pm 0.20 (\rm lumi.) \rm ~nb \nonumber 
\end{equation}
}
\end{itemize}

The corresponding predictions from HVQDIS are:

\begin{eqnarray}
\sigma(D^{+}) &=& 4.42 ~ ^{+0.86}_{-0.62} ~(\rm syst.) ~^{+0.42}_{-0.60} ~(\rm had.) \rm ~nb \nonumber \\ 
\sigma^{\rm antitag}(D^{0}) &=&  9.25 ~ ^{+1.79}_{-1.29} ~(\rm syst.)  ~^{+0.52}_{-0.96} ~(\rm had.) \rm ~nb \nonumber 
\end{eqnarray}

where ``had.'' and ``br.'' represent the uncertainty on the HVQDIS prediction due to the uncertainties of the hadronisation fraction $f(c\rightarrow D)$ and decay-chain branching ratios, respectively. The predictions used the default parameter settings as discussed in Section \ref{SEC:THEORY}. The quadratic sum of the other uncertainties of these predictions is shown with the ``syst.'' label. A small contribution ($\sim 2\%$) to the total cross sections arises from $D$ mesons produced in $b\bar{b}$ events. All predictions include a $b\bar{b}$ contribution calculated in each bin with HVQDIS. The HVQDIS predictions are in agreement with the data.

\begin{figure}
\begin{center}
\includegraphics[width=0.4\textwidth]{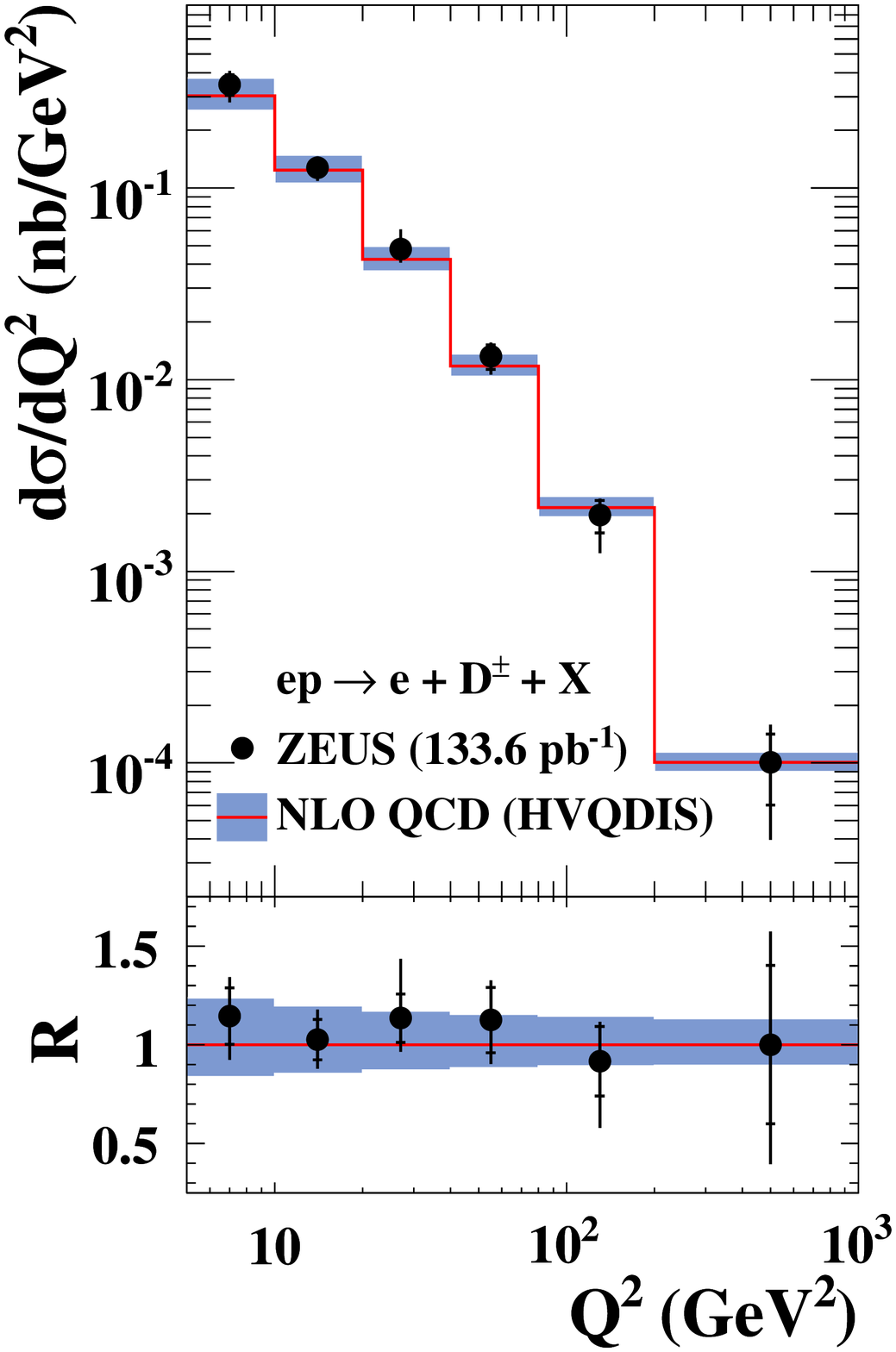}
\put(-48,255){\makebox(0,0)[tl]{\large (a)}}
\includegraphics[width=0.4\textwidth]{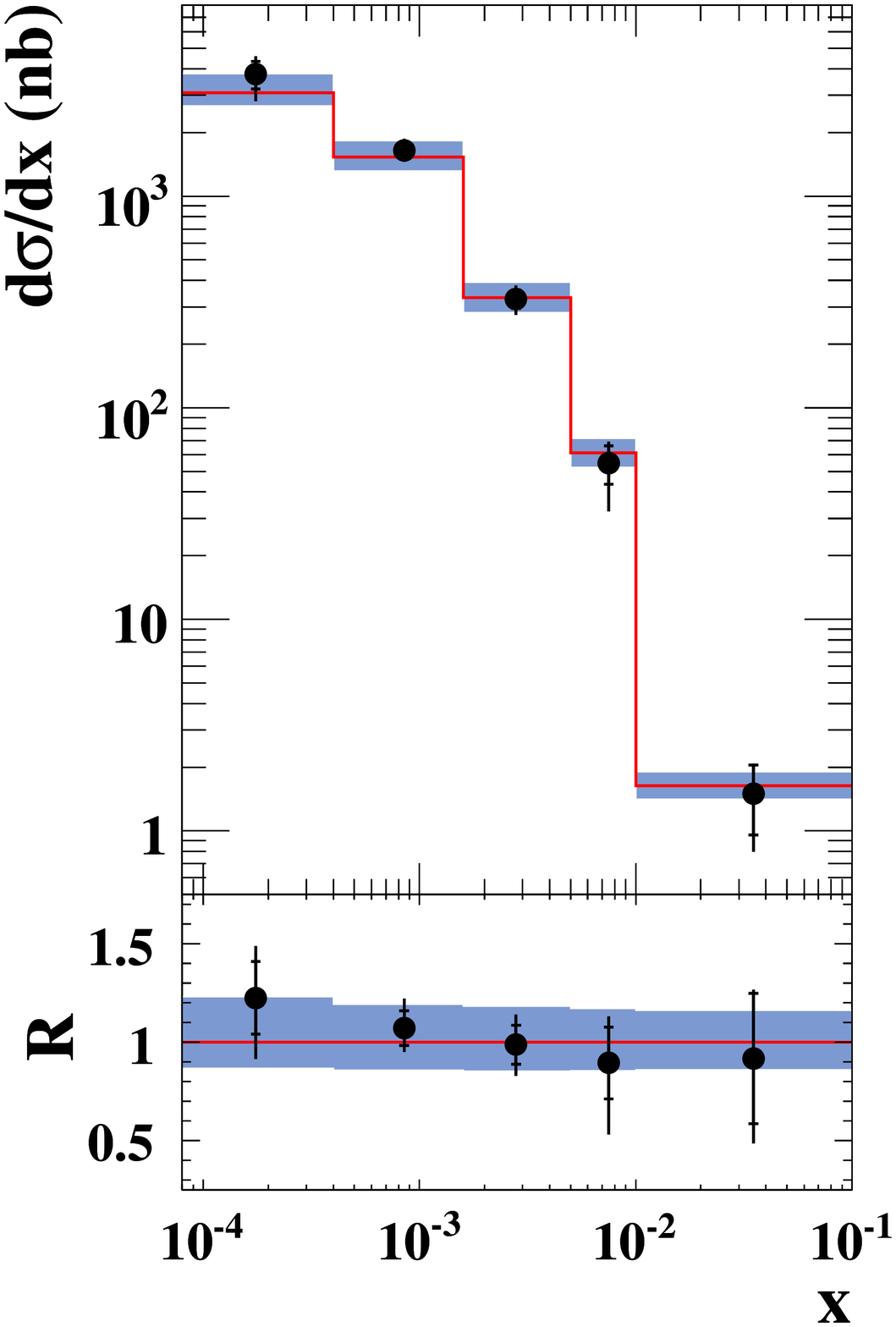}
\put(-48,255){\makebox(0,0)[tl]{\large (b)}} 
\put(-210, 285){\makebox(0,0)[tl]{\bf \Huge ZEUS}} \\
\includegraphics[width=0.4\textwidth]{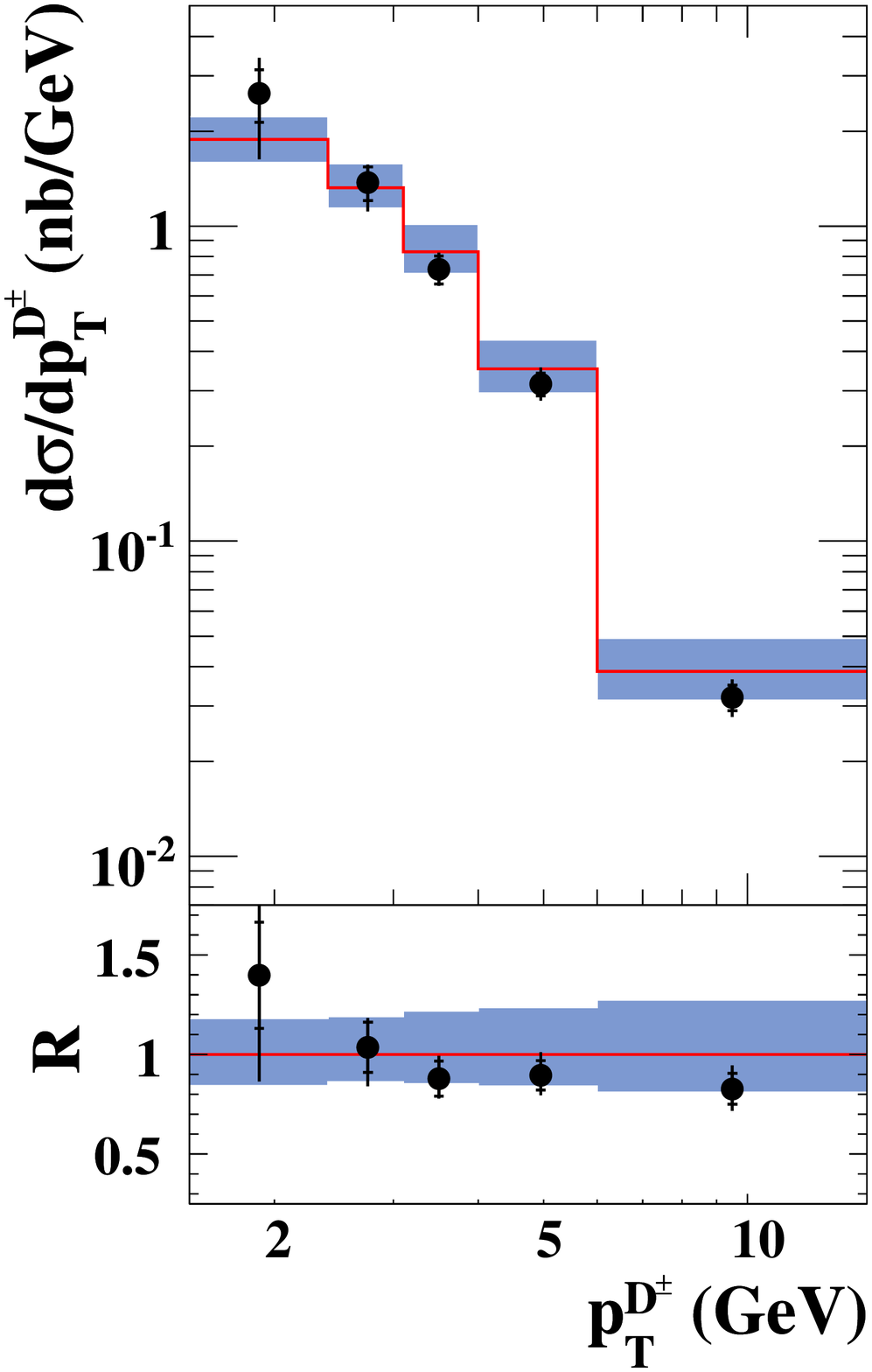}
\put(-48,255){\makebox(0,0)[tl]{\large (c)}}
\includegraphics[width=0.4\textwidth]{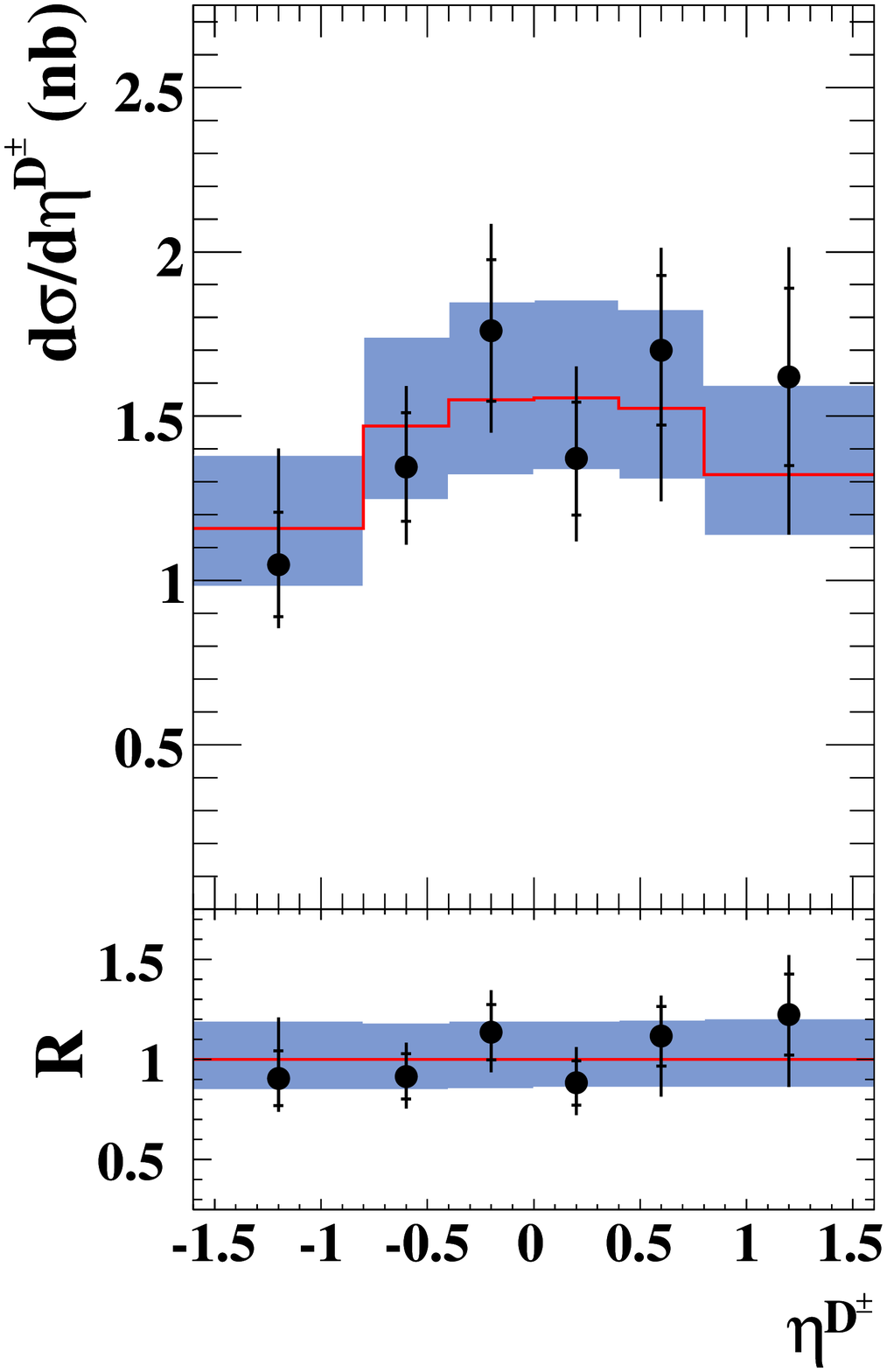}
\put(-48,255){\makebox(0,0)[tl]{\large (d)}}
\caption{Differential cross sections for $D^{\pm}$ mesons as a function of (a) $Q^{2}$, (b) $x$, (c) $p_{T}^{D^{\pm}}$ and (d) $\eta^{D^{\pm}}$ compared to the NLO QCD predictions of HVQDIS. Statistical uncertainties are shown by the inner error bars. Statistical and systematic uncertainties added in quadrature are shown by the outer error bars with the shaded region representing the uncertainty of the HVQDIS prediction. The ratios, R, of the cross sections to the central HVQDIS prediction are also shown in the lower section of each plot.}
\label{FIG:dplus_singdiff} 
\end{center}
\end{figure}

\begin{figure}
\begin{center}
\includegraphics[width=0.4\textwidth]{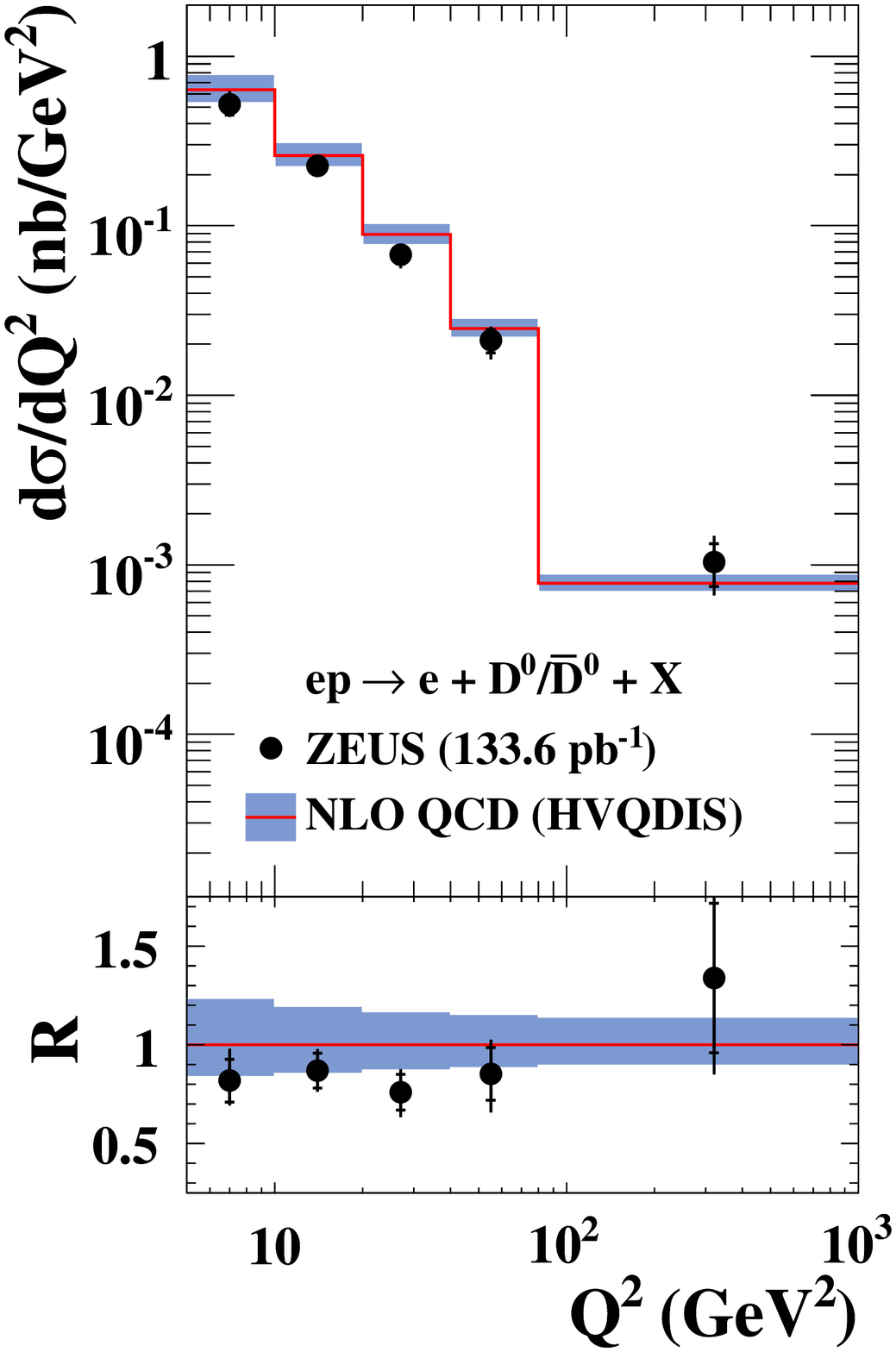}
\put(-48,255){\makebox(0,0)[tl]{\large (a)}}
\includegraphics[width=0.4\textwidth]{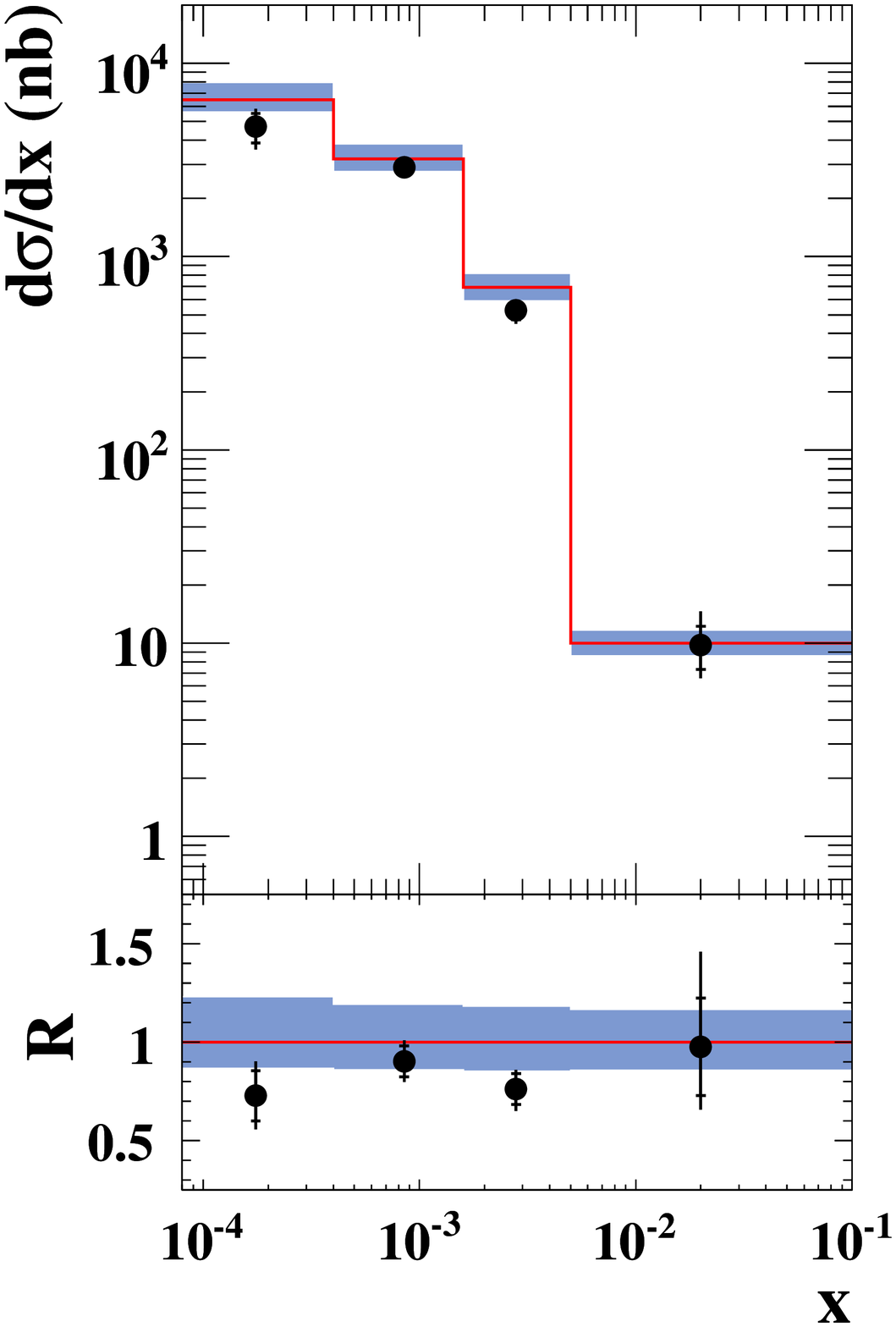}
\put(-48,255){\makebox(0,0)[tl]{\large (b)}} 
\put(-210, 285){\makebox(0,0)[tl]{\bf \Huge ZEUS}} \\
\includegraphics[width=0.4\textwidth]{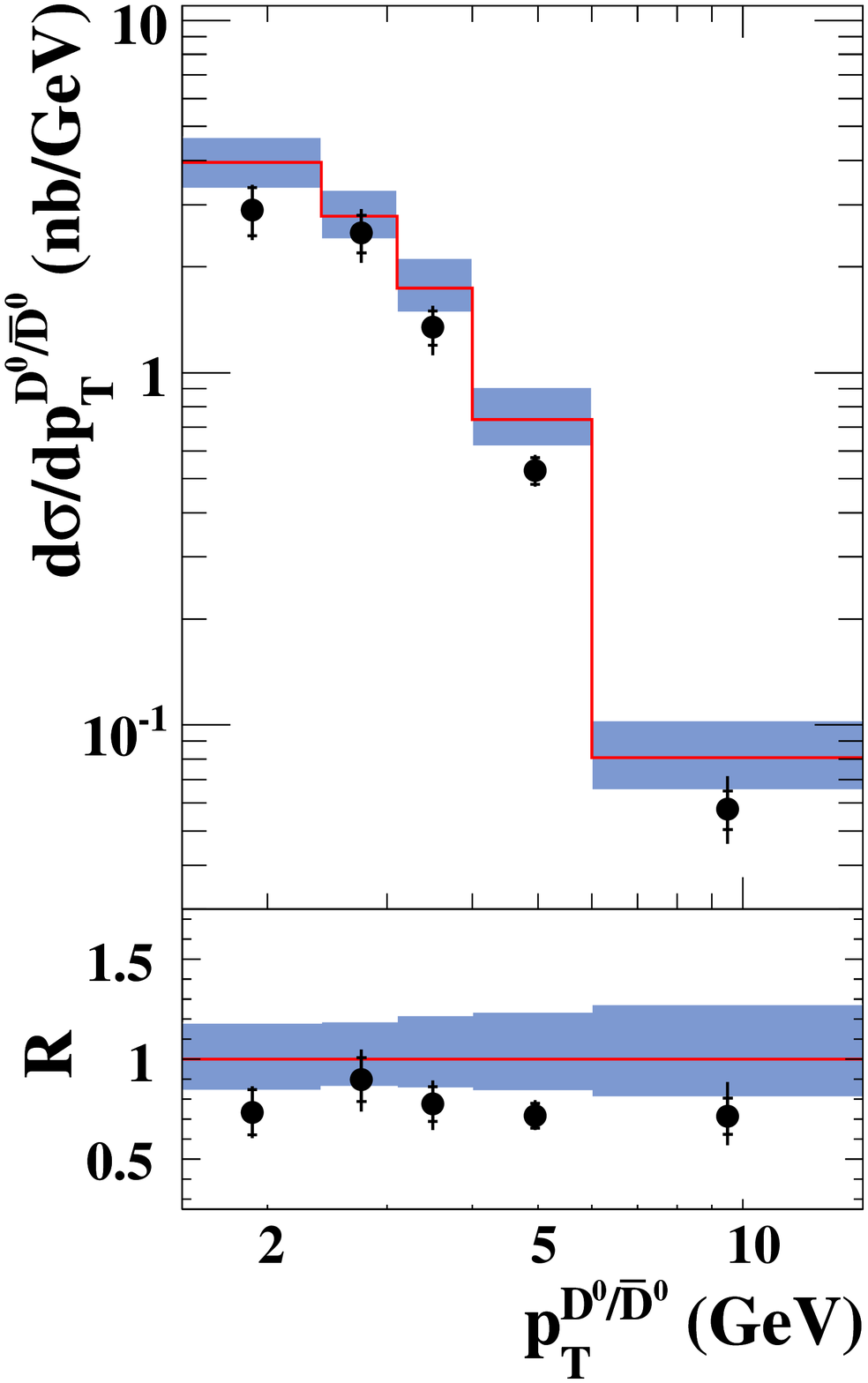}
\put(-48,255){\makebox(0,0)[tl]{\large (c)}}
\includegraphics[width=0.4\textwidth]{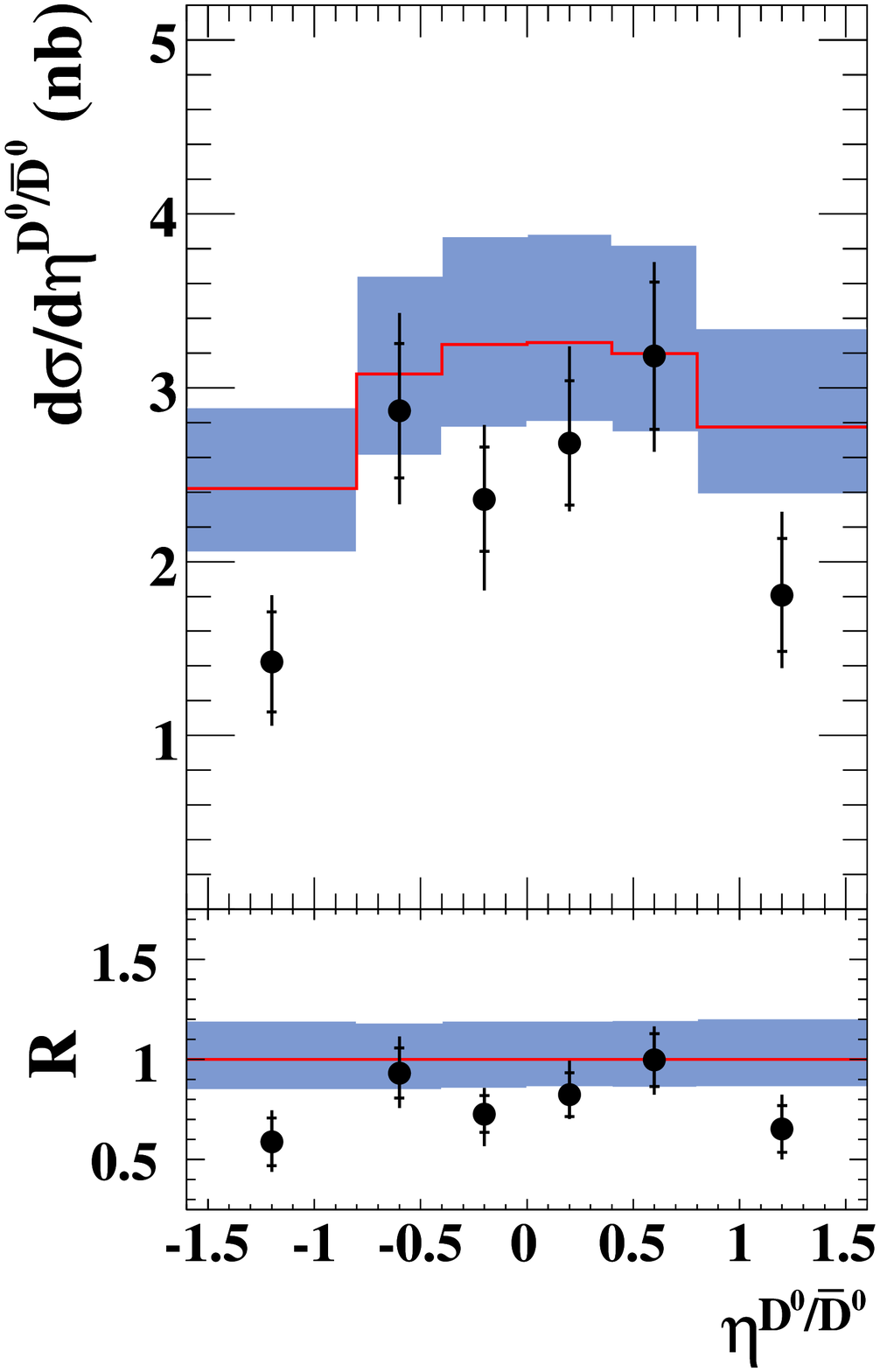}
\put(-48,255){\makebox(0,0)[tl]{\large (d)}}
\caption{Differential cross sections for $D^{0}/\bar{D}^{0}$ mesons not from $D^{*\pm}$ decay as a function of (a) $Q^{2}$, (b) $x$, (c) $p_{T}^{D^{0}/\bar{D}^{0}}$ and (d) $\eta^{D^{0}/\bar{D}^{0}}$ compared to the NLO QCD predictions of HVQDIS. Statistical uncertainties are shown by the inner error bars. Statistical and systematic uncertainties added in quadrature are shown by the outer error bars with the shaded region representing the uncertainty of the HVQDIS prediction. The ratios, R, of the cross sections to the central HVQDIS prediction are also shown in the lower section of each plot.}
\label{FIG:dzero_singdiff} 
\end{center}
\end{figure}

The differential $D^{+}$ and $D^{0}$ cross sections as functions of $Q^{2}, x, p_{T}^{D}$ and  $\eta^{D}$ are shown in Figs. \ref{FIG:dplus_singdiff} and \ref{FIG:dzero_singdiff} and given in Tables \ref{tab:dplus_single} and \ref{tab:dzero_single}. The cross sections in $Q^{2}$ and $x$ both fall by about three orders of magnitude in the measured region. The cross section in $p_{T}^{D}$ falls by about two orders of magnitude and there is no significant dependence on $\eta^{D}$. The HVQDIS predictions describe the shape of all measured differential cross sections well. The slight difference in normalisation in Fig.~\ref{FIG:dzero_singdiff} reflects the difference of the corresponding total cross section. 

\boldmath
\section{Extraction of  $F_{2}^{c\bar{c}}$}
\unboldmath
 
The open-charm contribution, $F_{2}^{c\bar{c}}$, to the proton structure function, $F_{2}$, can be defined in terms of the inclusive double-differential $c\bar{c}$ cross section in $x$ and $Q^{2}$ by
 
\begin{equation}
\frac{d^{2}\sigma^{c\bar{c}}(x, Q^{2})} {dxdQ^{2}} = \frac{2\pi \alpha^{2}} {xQ^{4}} \left\{ \left[ 1+ \left(1-y \right)^{2}  \right]  F_{2}^{c\bar{c}}(x,Q^{2}) - y^{2}F_{L}^{c\bar{c}}(x,Q^{2}) \right\}.
\end{equation}

In this paper, the $c\bar{c}$ cross section is obtained by measuring the $D^{+}$ and 
$D^{0}$ production cross sections and employing the hadronisation fraction 
$f(c \rightarrow D)$ to derive the total charm cross section. A limited kinematic region 
is accessible for the measurement of $D$ mesons, therefore a prescription for 
extrapolating to the full kinematic phase space is needed. The measured value of 
$F_{2}^{c\bar{c}}$ in a bin $i$ is calculated with

\begin{equation}
F_{2, \rm meas}^{c\bar{c}}(x_{i}, Q^{2}_{i}) = \frac{\sigma_{i, \rm meas}(ep\rightarrow DX)}{\sigma_{i, \rm theo}(ep \rightarrow DX)}F_{2, \rm theo}^{c\bar{c}}(x_{i}, Q^{2}_{i}),
\end{equation}
where $\sigma_{i, \rm meas}$ is the cross section in the bin $i$ in the measured region of $p_{T}^{D}$ and $\eta^{D}$ and $\sigma_{i,\rm theo}$ is the corresponding cross section evaluated with HVQDIS. The value of  $F_{2, \rm theo}^{c\bar{c}}$ was calculated in FFNS from the NLO coefficient functions \cite{pr:d67:012007} using the same values of parameters as in the calculation of $\sigma_{i,\rm theo}$. The cross sections $\sigma_{i, \rm meas}(ep \rightarrow DX)$ were measured in bins of $Q^{2}$ and $y$ (Table \ref{tab:q2y}) and $F_{2}^{c\bar{c}}$ is quoted at representative $Q^{2}$ and $x$ values near the centre-of-gravity for each bin (Table \ref{tab:f2cc}).

Beauty contributions were subtracted from the data using the predictions obtained from HVQDIS. The contribution to the total cross section from $F_{L}^{c\bar{c}}$ calculated using the ZEUS NLO fit was, on average, 1.3$\%$ and at most $4.7\%$ \cite{pr:d69:012004} and was taken into account, in  $\sigma_{i,\rm theo}$, in the extraction of $F_{2}^{c\bar{c}}$. The size of the contribution from $F_{L}$ was similar to that in other PDFs.

The factor to extrapolate from the measurement range to the full phase space was estimated using HVQDIS and was found to vary from $\approx 1.5$  at high $Q^{2}$ to $\approx 3.2$ at low $Q^{2}$. A complete list of the extrapolation factors is given in Table \ref{tab:f2cc}.

The following uncertainties associated with the method of extrapolation were evaluated with the average effect given in parentheses:

\begin{itemize}

\item{changing the charm mass by $\mp 0.15$\,GeV consistently in the HVQDIS calculation and in the calculation of $F_{2, \rm theo}^{c\bar{c}}$ ($\pm 2\%$). The largest effect was seen at low $x$ and low $Q^{2}$ $(^{+7}_{-5}\%)$;}

\item{using the upper and lower predictions given by the uncertainty in the ZEUS NLO PDF fit, propagated from the experimental uncertainties of the fitted data, to perform the extraction of $F_{2}^{c\bar{c}}$ ($< 1\%$);}

\item{changing the contribution of beauty events subtracted from the data by a factor 2 ($^{+1}_{-2}\%$). The largest effect was seen at low $x$ and high $Q^{2}$ ($^{+3}_{-7}\%$);}

\item{a Lund string model in {\sc Rapgap} was used as in previous analyses~\cite{pr:d69:012004,epj:c12:35,Chekanov:2007ch} rather than the Peterson function in HVQDIS ($\pm 7\%$). The largest effect was seen at high $x$ and low $Q^{2}$ ($\pm 14\%$).}

\end{itemize} 

\begin{figure}
\begin{center}
\includegraphics[width=\textwidth]{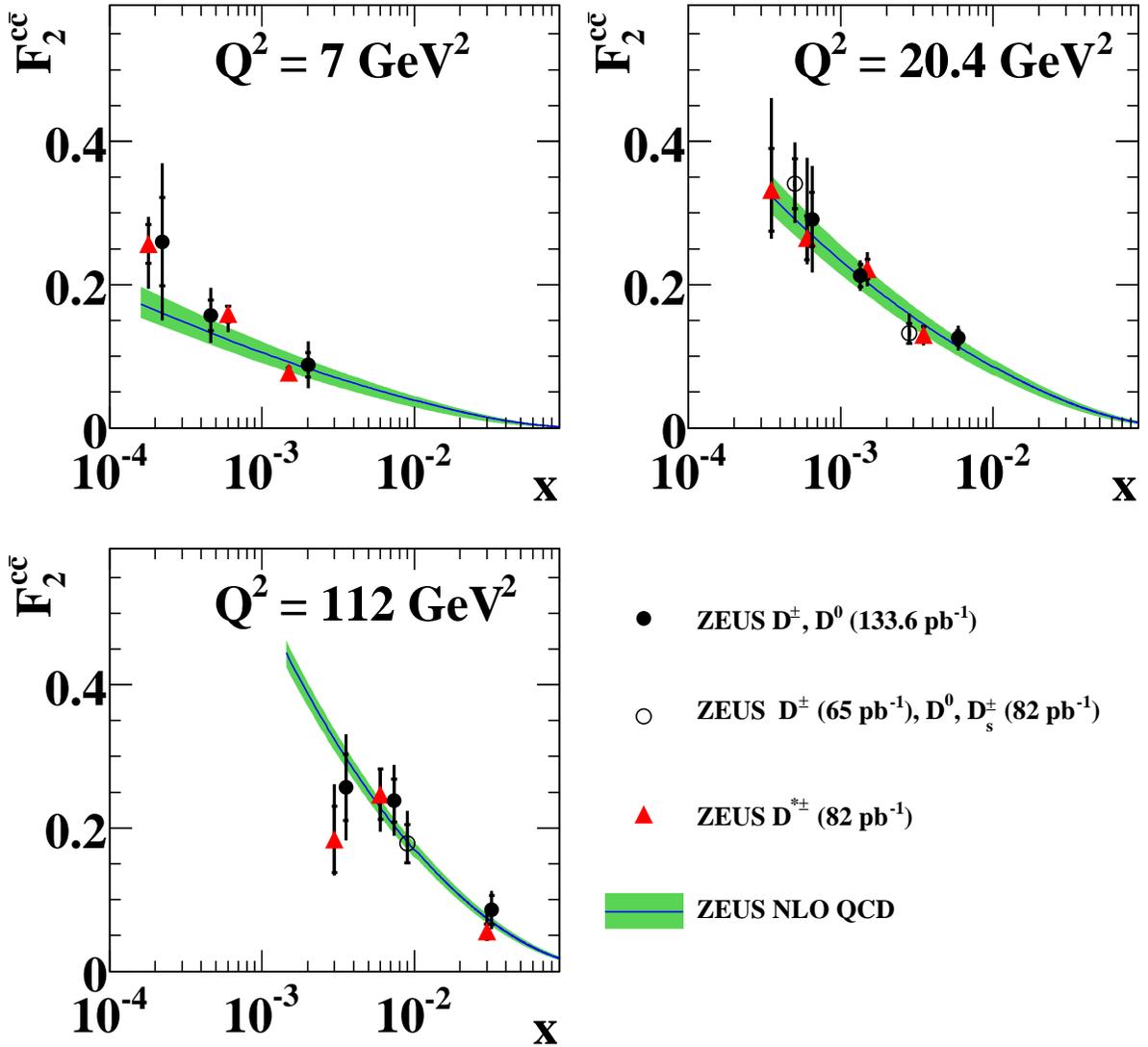}
\put(-250, 450){\makebox(0,0)[tl]{\bf \Huge ZEUS}}
\caption{Combined values of $F_{2}^{c\bar{c}}$ extracted from $D^{\pm}$ and $D^{0}/\bar{D}^{0}$ not from $D^{*\pm}$ (circles) as a function of $x$ in three bins of $Q^{2}$.  The data are shown with statistical uncertainties (inner bars) and statistical and systematic uncertainties added in quadrature (outer bars) and, where possible, are compared to previous ZEUS measurements with these mesons. The measurements have a further uncertainty of $3.3\%$ from the $D^{+} \rightarrow K^{-}\pi^{+}\pi^{+} (+c.c.)$ and  $D^{0} \rightarrow K^{-}\pi^{+} (+c.c.)$ branching ratios. The additional uncertainty from the luminosity mesaurements is $2.6\%$. The shaded band shows the predicted values of $F_{2}^{c\bar{c}}$ for values of $m_{c}$ between 1.35 and 1.65 GeV.}
\label{FIG:f2cc}
\end{center}
\end{figure}

The $F_{2}^{c\bar{c}}$ values measured from $D^{+}$ and $D^{0}$ data are combined using a procedure that accounts for the systematic and point-to-point correlations between the analyses \cite{proc:dis:2005:237}. The combined values of $F_{2}^{c\bar{c}}$ obtained from 
$D^{+}$ and $D^{0}$ production are given in Table~\ref{tab:f2cc_comb} and shown in Fig.~\ref{FIG:f2cc}. Also shown is the 
ZEUS NLO QCD fit which describes the data well for all $Q^{2}$ and $x$. The uncertainty 
of the theoretical prediction shown is that from uncertainty of the charm mass. Due to the improved statistical precision resulting 
from lifetime tags with the MVD, more measurements of $F_{2}^{c\bar{c}}$ were extracted than in the previous publication \cite{Chekanov:2007ch}. Also, extrapolation factors 
were significantly reduced, from e.g. a value of about 5--6 to about 2 at 
$Q^2=20.4$\,GeV$^2$, due to the extension of the kinematic range to lower $p_{T}^{D}$. At high $Q^{2}$ these results are competitive with $D^{*\pm}$ based measurements \cite{pr:d69:012004}.


\section{Conclusions}

The production of the charm mesons $D^{+}$ and $D^{0}$ has been measured with the ZEUS detector in the kinematic range $5 < Q^{2} < 1000$\,GeV$^{2}$, $0.02 < y < 0.7$, $1.5 < p_{T}^{D} < 15$\,GeV and $|\eta^{D}| < 1.6$. Combinatorial background to the $D$ meson signals was reduced by using the ZEUS microvertex detector to reconstruct displaced secondary vertices.

The measured $D$ meson cross sections were compared to the predictions of NLO QCD with the proton PDFs extracted from inclusive DIS data. A good description was found.

The visible cross sections in bins of $y$ and $Q^{2}$ were used to extract the open-charm contribution, $F_{2}^{c\bar{c}}$, to the proton structure function, $F_{2}$. The extraction used factors calculated within the framework of NLO QCD.

The use of the microvertex detector has increased the precision and allowed an extension in the kinematic range to lower values of $p_{T}^{D}$  compared to previous results. Along with previous measurements of $F_{2}^{c\bar{c}}$, the results presented here provide a direct constraint on the gluon density of the proton.

\section{Acknowledgements}

The strong support and encouragement of the DESY Directorate has been invaluable, and we are much indebted to the HERA machine group for their inventiveness and diligent efforts. The design, construction and installation of the ZEUS detector were made possible by the ingenuity and dedicated efforts of many people from inside DESY and from the home institutes who are not listed as authors. Their contributions are acknowledged with great appreciation.

{
\def\bibname{\Large\bf References}
\def\refname{\Large\bf References}
\pagestyle{plain}
\ifzeusbst
  \bibliographystyle{./BiBTeX/bst/l4z_default}
\fi
\ifzdrftbst
  \bibliographystyle{./BiBTeX/bst/l4z_draft}
\fi
\ifzbstepj
  \bibliographystyle{./BiBTeX/bst/l4z_epj}
\fi
\ifzbstnp
  \bibliographystyle{./BiBTeX/bst/l4z_np}
\fi
\ifzbstpl
  \bibliographystyle{./BiBTeX/bst/l4z_pl}
\fi
{\raggedright
\bibliography{./BiBTeX/user/syn.bib,%
              ./BiBTeX/bib/l4z_articles.bib,%
              ./BiBTeX/bib/l4z_books.bib,%
              ./BiBTeX/bib/l4z_conferences.bib,%
              ./BiBTeX/bib/l4z_h1.bib,%
              ./BiBTeX/bib/l4z_misc.bib,%
              ./BiBTeX/bib/l4z_old.bib,%
              ./BiBTeX/bib/l4z_preprints.bib,%
              ./BiBTeX/bib/l4z_replaced.bib,%
              ./BiBTeX/bib/l4z_temporary.bib,%
              ./BiBTeX/bib/l4z_zeus.bib}}
}
\vfill\eject

\begin{table}
\begin{center}
\begin{tabular}[width = \textwidth]{|r@{, }l|r@{.}l@{    }r@{.}l@{    }r@{.}l@{    }r@{.}l|}
\hline
\multicolumn{2}{|c|}{$Q^{2}$ bin}  & \multicolumn{2}{c}{$d\sigma/dQ^{2}$} & \multicolumn{2}{c}{$\Delta_{\rm stat}$}& \multicolumn{4}{c|}{$\Delta_{\rm syst}$} \\
\multicolumn{2}{|c|}{(GeV$^{2}$)}  & \multicolumn{8}{c|}{(nb/GeV$^{2}$)} \\
\hline
\hline
5 & 10 & 0&35 & $\pm$0&04 & +0&04 & $-$0&05 \\
10 & 20 & 0&13 & $\pm$0&01 & +0&01 & $-$0&01 \\
20 & 40 & 0&048 & $\pm$0&005 & +0&012 & $-$0&005  \\
40 & 80 & 0&013 & $\pm$0&002 & +0&001 & $-$0&002  \\
80 & 200 & 0&0020 & $\pm$0&0004 & +0&0002 & $-$0&0006  \\
200 & 1000 & 0&00010 & $\pm$0&00004 & +0&00004 & $-$0&00005  \\
 \hline  
\hline
\multicolumn{2}{|c|}{$x$ bin}  & \multicolumn{2}{c}{$d\sigma/dx$} & \multicolumn{2}{c}{$\Delta_{\rm stat}$}& \multicolumn{4}{c|}{$\Delta_{\rm syst}$}\\
 \multicolumn{2}{|c|}{}& \multicolumn{8}{c|}{(nb)} \\
\hline
\hline
0.00008 & 0.0004 & 3773&0 & $\pm$566&0 & +577&0 & $-$773&0 \\
0.00040 & 0.0016 & 1643&0 & $\pm$136&0 & +183&0 & $-$128&0 \\
0.0016 & 0.005 & 327&0 & $\pm$33&0 & +39&0 & $-$42&0 \\
0.005 & 0.01 & 55&0 & $\pm$11&0 & +9&0 & $-$19&0 \\
0.01 & 0.1 & 1&5 & $\pm$0&5 & +0&2 & $-$0&5 \\
 \hline  
\hline
\multicolumn{2}{|c|}{$p_{T}^{D^{\pm}}$ bin}  & \multicolumn{2}{c}{$d\sigma/dp_{T}^{D^{\pm}}$} & \multicolumn{2}{c}{$\Delta_{\rm stat}$}& \multicolumn{4}{c|}{$\Delta_{\rm syst}$} \\
 \multicolumn{2}{|c|}{(GeV)}& \multicolumn{8}{c|}{(nb/GeV)} \\
\hline
\hline
1.5 & 2.4 & 2&63 & $\pm$0&50 & +0&59 & $-$0&87 \\
2.4 & 3.1 & 1&37 & $\pm$0&17 & +0&10 & $-$0&20 \\
3.1 & 4.0 & 0&73 & $\pm$0&07 & +0&06 & $-$0&04 \\
4.0 & 6.0 & 0&32 & $\pm$0&03 & +0&03 & $-$0&02 \\
6.0 & 15.0 & 0&032 & $\pm$0&003 & +0&003 & $-$0&003 \\
 \hline  
\hline
\multicolumn{2}{|c|}{$\eta^{D^{\pm}}$ bin}  & \multicolumn{2}{c}{$d\sigma/d\eta^{D^{\pm}}$} & \multicolumn{2}{c}{$\Delta_{\rm stat}$}& \multicolumn{4}{c|}{$\Delta_{\rm syst}$}\\
 \multicolumn{2}{|c|}{}& \multicolumn{8}{c|}{(nb)} \\
\hline
\hline
$-$1.6 & $-$0.8 & 1&05 & $\pm$0&16 & +0&32 & $-$0&11 \\
$-$0.8 & $-$0.4 & 1&35 & $\pm$0&17 & +0&18 & $-$0&17 \\
$-$0.4 & 0.0 & 1&76 & $\pm$0&22 & +0&24 & $-$0&22 \\
0.0 & 0.4 & 1&37 & $\pm$0&17 & +0&22 & $-$0&18 \\
0.4 & 0.8 & 1&70 & $\pm$0&23 & +0&21 & $-$0&40 \\
0.8 & 1.6 & 1&62 & $\pm$0&27 & +0&29 & $-$0&40 \\
 \hline  
\end{tabular}
\caption{Measured $D^{\pm}$ cross sections as a function of $Q^{2}$, $x$, $p_{T}^{D^{\pm}}$ and $\eta^{D^{\pm}}$ for $5 < Q^{2} < 1000$ GeV$^{2}$, $0.02 < y < 0.7$, $1.5 < p_{T}^{D^{\pm}} < 15$ GeV and $|\eta^{D^{\pm}}| < 1.6$. The statistical and systematic uncertainties are shown separately. The cross sections have further uncertainties of $3.5\%$ from the $D^{+} \rightarrow K^{-}\pi^{+}\pi^{+} (+c.c.)$ branching ratio, and $2.6\%$ from the uncertainty in the luminosity measurement.}
\label{tab:dplus_single}
\end{center}
\end{table}
\begin{table}
\begin{center}
\begin{tabular}[width = \textwidth]{|r@{, }l|r@{.}l@{    }r@{.}l@{    }r@{.}l@{    }r@{.}l|}
\hline
\multicolumn{2}{|c|}{$Q^{2}$ bin}  & \multicolumn{2}{c}{$d\sigma/dQ^{2}$} & \multicolumn{2}{c}{$\Delta_{\rm stat}$}& \multicolumn{4}{c|}{$\Delta_{\rm syst}$} \\
\multicolumn{2}{|c|}{(GeV$^{2}$)}  & \multicolumn{8}{c|}{(nb/GeV$^{2}$)} \\
\hline
\hline
5 & 10 & 0&52 & $\pm$0&07 & +0&08 & $-$0&04 \\
10 & 20 & 0&23 & $\pm$0&02 & +0&02 & $-$0&02 \\
20 & 40 & 0&067 & $\pm$0&008 & +0&007 & $-$0&008 \\
40 & 80 & 0&021 & $\pm$0&003 & +0&003 & $-$0&003 \\
80 & 1000 & 0&0010 & $\pm$0&0003 &+0&0003 & $-$0&0002 \\
 \hline  
\hline
\multicolumn{2}{|c|}{$x$ bin}  & \multicolumn{2}{c}{$d\sigma/dx$} & \multicolumn{2}{c}{$\Delta_{\rm stat}$}& \multicolumn{4}{c|}{$\Delta_{\rm syst}$} \\
\multicolumn{2}{|c|}{}  & \multicolumn{8}{c|}{(nb)} \\
\hline
\hline
0.00008 & 0.0004 & 4697&0 & $\pm$824&0 & +769&0 & $-$743&0 \\
0.00040 & 0.0016 & 2896&0 & $\pm$254&0 & +235&0 & $-$225&0 \\
0.0016 & 0.005 & 527&0 & $\pm$54&0 & +41&0 & $-$55&0 \\
0.005 & 0.1 & 10&0 & $\pm$2&0 & +4&0 & $-$2&0 \\
 \hline  
\hline
\multicolumn{2}{|c|}{$p_{T}^{D^{0}/\bar{D}^{0}}$ bin}  & \multicolumn{2}{c}{$d\sigma/dp_{T}^{D^{0}/\bar{D}^{0}}$} & \multicolumn{2}{c}{$\Delta_{\rm stat}$}& \multicolumn{4}{c|}{$\Delta_{\rm syst}$} \\
\multicolumn{2}{|c|}{(GeV)}  & \multicolumn{8}{c|}{(nb/GeV)} \\
\hline
\hline
1.5 & 2.4 & 2&90 & $\pm$0&45 & +0&26 & $-$0&26 \\
2.4 & 3.1 & 2&49 & $\pm$0&31 & +0&29 & $-$0&32 \\
3.1 & 4.0 & 1&35 & $\pm$0&15 & +0&14 & $-$0&17 \\
4.0 & 6.0 & 0&53 & $\pm$0&05 & +0&03 & $-$0&02 \\
6.0 & 15.0 & 0&058 & $\pm$0&007 & +0&012 & $-$0&009 \\
 \hline  
\hline
\multicolumn{2}{|c|}{$\eta^{D^{0}/\bar{D}^{0}}$ bin } & \multicolumn{2}{c}{$d\sigma/d\eta^{D^{0}/\bar{D}^{0}}$} & \multicolumn{2}{c}{$\Delta_{\rm stat}$}& \multicolumn{4}{c|}{$\Delta_{\rm syst}$} \\
\multicolumn{2}{|c|}{}  & \multicolumn{8}{c|}{(nb)} \\
\hline
\hline
$-$1.6 & $-$0.8 & 1&42 & $\pm$0&29 & +0&25 & $-$0&23 \\
$-$0.8 & $-$0.4 & 2&87 & $\pm$0&39 & +0&41 & $-$0&37 \\
$-$0.4 & 0.0 & 2&36 & $\pm$0&30 & +0&30 & $-$0&43 \\
0.0 & 0.4 & 2&68 & $\pm$0&36 & +0&42 & $-$0&16 \\
0.4 & 0.8 & 3&18 & $\pm$0&42 & +0&34 & $-$0&36 \\
0.8 & 1.6 & 1&81 & $\pm$0&33 & +0&35 & $-$0&27 \\
 \hline  
\end{tabular}
\caption{Measured cross sections for $D^{0}/\bar{D}^{0}$ not coming from a $D^{*\pm}$ as a function of $Q^{2}$, $x$, $p_{T}^{D^{0}/\bar{D}^{0}}$ and $\eta^{D^{0}/\bar{D}^{0}}$  for $5 < Q^{2} < 1000$ GeV$^{2}$, $0.02 < y < 0.7$, $1.5 < p_{T}^{D^{0}/\bar{D}^{0}} < 15$ GeV and $|\eta^{D^{0}/\bar{D}^{0}}| < 1.6$. The statistical and systematic uncertainties are shown separately. The cross sections have further uncertainties of $1.9\%$ from the $D^{0} \rightarrow K^{-}\pi^{+} (+c.c.)$ branching ratio, and $2.6\%$ from the uncertainty in the luminosity measurement.}
\label{tab:dzero_single}
\end{center}
\end{table}


\begin{table}
\begin{center}
\begin{tabular}{|c|c|cccc|}
\hline
 $Q^{2}$ bin   & $y$ bin & $\sigma(D^{\pm})$ & $\Delta_{\rm stat}$ &  \multicolumn{2}{c|}{$\Delta_{\rm syst}$} \\
 (GeV$^{2}$)  & &\multicolumn{4}{c|}{(nb)} \\
\hline
\hline
& 0.02, 0.12 & 0.52 & $\pm$0.13 & +0.17 & $-$0.14\\
5, 9 & 0.12, 0.30 & 0.59 & $\pm$0.11 & +0.08 & $-$0.17\\
& 0.30, 0.70 & 0.56 & $\pm$0.17 & +0.17 & $-$0.14\\
\hline
& 0.02, 0.12 & 0.94 & $\pm$0.10 & +0.07 & $-$0.13\\
9, 44 & 0.12, 0.30 & 0.96 & $\pm$0.09 & +0.06 & $-$0.06\\
& 0.30, 0.70 & 0.73 & $\pm$0.12 & +0.08 & $-$0.20\\
\hline
& 0.02, 0.12 & 0.20 & $\pm$0.05 & +0.01 & $-$0.03\\
44, 1000 & 0.12, 0.30 & 0.35 & $\pm$0.06 & +0.05 & $-$0.08\\
& 0.30, 0.70 & 0.24 & $\pm$0.05 & +0.03 & $-$0.06\\
\hline
\multicolumn{6}{c}{ } \\
\multicolumn{6}{c}{ } \\
\multicolumn{6}{c}{ } \\
\multicolumn{6}{c}{ } \\
\hline
$Q^{2}$ bin  & $y$ bin & $\sigma(D^{0}/\bar{D}^{0})$ & $\Delta_{\rm stat}$ &  \multicolumn{2}{c|}{$\Delta_{\rm syst}$}\\
(GeV$^{2}$)& & \multicolumn{4}{c|}{(nb)} \\
\hline
\hline
& 0.02, 0.12 & 0.80 & $\pm$0.24 & +0.23 & $-$0.16\\
5, 9 & 0.12, 0.30 & 0.92 & $\pm$0.20 & +0.13 & $-$0.12\\
& 0.30, 0.70 & 0.48 & $\pm$0.17 & +0.11 & $-$0.14\\
\hline
& 0.02, 0.12 & 1.62 & $\pm$0.18 & +0.10 & $-$0.13\\
9, 44 & 0.12, 0.30 & 1.42 & $\pm$0.15 & +0.05 & $-$0.06\\
& 0.30, 0.70 & 1.22 & $\pm$0.24 & +0.24 & $-$0.18\\
\hline
& 0.02, 0.12 & 0.19 & $\pm$0.09 & +0.06 & $-$0.04\\
44, 1000 & 0.12, 0.30 & 0.54 & $\pm$0.09 & +0.06 & $-$0.04\\
& 0.30, 0.70 & 0.54 & $\pm$0.15 & +0.15 & $-$0.18\\
\hline
\end{tabular}
\put(-170, 268){\makebox(0,0)[tl]{\bf \Huge $D^{\pm}$}} 
\put(-190, 0){\makebox(0,0)[tl]{\bf \Huge $D^{0}/\bar{D}^{0}$}} 
\caption{Measured cross sections for $D^{\pm}$ and $D^{0}/\bar{D}^{0}$ not coming from a $D^{*\pm}$ in each of the $Q^{2}$ and $y$ bins for $5 < Q^{2} < 1000$ GeV$^{2}$, $0.02 < y < 0.7$, $1.5 < p_{T}^{D} < 15$ GeV and $|\eta^{D}| < 1.6$. The statistical and systematic uncertainties are shown separately. The $D^{\pm}$ and $D^{0}/\bar{D}^{0}$ cross sections have further uncertainties of $3.5\%$ and $1.9\%$ from the $D^{+}\rightarrow K^{-}\pi^{+}\pi^{+} (+c.c.)$ and  $D^{0}\rightarrow K^{-}\pi^{+} (+c.c.)$ branching ratios. The additional uncertainty from the luminosity measurements is $2.6\%$.}
\label{tab:q2y}
\end{center}
\end{table}


\begin{table}
\begin{center}
\begin{tabular}{|c|c|cccc|ccc|}
\hline
$Q^{2}$ & $x$ & $F_{2}^{c\bar{c}}$ & $\Delta_{\rm stat}$ &  \multicolumn{2}{c|}{$\Delta_{\rm syst}$} & \multicolumn{2}{c}{$\Delta_{\rm extrap}$} & factor\\
(GeV$^{2}$) & & & & & & & &  \\
\hline
\hline
 & 0.00022 & 0.295 & $\pm$0.092 & +0.091 & $-$0.074 &  +0.026 & $-$0.022 & 3.2\\
7.0 & 0.00046 & 0.176 & $\pm$0.031 & +0.023 & $-$0.050 & +0.010 & $-$0.008 & 2.3\\
 & 0.00202 & 0.091 & $\pm$0.023 & +0.030 & $-$0.025 &  +0.013 & $-$0.014 & 3.1\\
\hline
 & 0.00065 & 0.319 & $\pm$0.054 & +0.037 & $-$0.086 &  +0.022 & $-$0.020 & 2.5\\
20.4 & 0.00134 & 0.241 & $\pm$0.024 & +0.016 & $-$0.014 & +0.013 & $-$0.013 & 1.8\\
 & 0.00588 & 0.131 & $\pm$0.015 & +0.010 & $-$0.018 &  +0.009 & $-$0.009 & 2.4\\
\hline
 & 0.00356 & 0.260 & $\pm$0.058 & +0.029 & $-$0.066 &  +0.020 & $-$0.025 & 1.7\\
112.0 & 0.00738 & 0.280 & $\pm$0.049 & +0.038 & $-$0.064 & +0.032 & $-$0.033 & 1.5\\
 & 0.03230 & 0.089 & $\pm$0.024 & +0.004 & $-$0.015 &  +0.002 & $-$0.002 & 2.4\\
\hline
\multicolumn{7}{c}{  } \\
\multicolumn{7}{c}{  } \\
\multicolumn{7}{c}{  } \\
\multicolumn{7}{c}{  } \\
\hline
$Q^{2}$ & $x$ & $F_{2}^{c\bar{c}}$ & $\Delta_{\rm stat}$ &  \multicolumn{2}{c|}{$\Delta_{\rm syst}$} & \multicolumn{2}{c}{$\Delta_{\rm extrap}$} & factor\\
(GeV$^{2}$) & & & & & & & &  \\
\hline
\hline
 & 0.00022 & 0.116 & $\pm$0.042 & +0.028 & $-$0.035 & +0.010 & $-$0.009 & 3.2\\
7.0 & 0.00046 & 0.131 & $\pm$0.029 & +0.019 & $-$0.017 & +0.007 & $-$0.006 & 2.3\\
 & 0.00202 & 0.068 & $\pm$0.020 & +0.019 & $-$0.014 & +0.010 & $-$0.010 & 3.1\\
\hline
 & 0.00065 & 0.252 & $\pm$0.051 & +0.049 & $-$0.037 & +0.017 & $-$0.016 & 2.5\\
20.4 & 0.00134 & 0.169 & $\pm$0.019 & +0.006 & $-$0.007 & +0.009 & $-$0.009 & 1.8\\
 & 0.00588 & 0.109 & $\pm$0.012 & +0.006 & $-$0.009 & +0.007 & $-$0.008 & 2.4\\
\hline
 & 0.00356 & 0.280 & $\pm$0.086 & +0.077 & $-$0.096 & +0.022 & $-$0.027 & 1.7\\
112.0 & 0.00738 & 0.203 & $\pm$0.037 & +0.024 & $-$0.016 & +0.023 & $-$0.024 & 1.5\\
 & 0.03230 & 0.040 & $\pm$0.019 & +0.012 & $-$0.008 & +0.001 & $-$0.001 & 2.4\\
\hline
\end{tabular}
\put(-210, 270){\makebox(0,0)[tl]{\bf \Huge $D^{\pm}$}} 
\put(-230, 0){\makebox(0,0)[tl]{\bf \Huge $D^{0}/\bar{D}^{0}$}} 
\caption{The extracted values of $F_{2}^{c\bar{c}}$ from the production cross sections of $D^{\pm}$ and $D^{0}/\bar{D}^{0}$ not coming from $D^{*\pm}$ at each $Q^{2}$ and $x$ value. The statistical, systematic and extrapolation uncertainties are shown separately. The values of the extrapolation factor used to correct to the full $p_{T}^{D}$ and $\eta^{D}$ phase space are also shown. The values extracted from $D^{\pm}$ and $D^{0}/\bar{D}^{0}$ have further uncertainties as detailed in the caption to Table~\ref{tab:q2y}.}
\label{tab:f2cc}
\end{center}
\end{table}
\begin{table}
\begin{center}
\begin{tabular}{|c|c|ccc|cc|}
\hline
$Q^{2}$ & $x$ & $F_{2}^{c\bar{c}}$ & $\Delta_{\rm stat}$ & $\Delta_{\rm syst}$ & \multicolumn{2}{c|}{$\Delta_{\rm extrap}$}\\
 (GeV$^{2}$) & & & & & & \\
\hline
\hline
 & 0.00022 & 0.260 & $\pm$0.062 & $\pm0.091$ & +0.007 & $-$0.067\\
7.0 & 0.00046 & 0.157 & $\pm$0.022 &  $\pm0.031$ & +0.016 & $-$0.035\\
 & 0.00202 & 0.088 & $\pm$0.017 & $\pm0.028$ & +0.009 & $-$0.016\\
\hline
 & 0.00065 & 0.291 & $\pm$0.038 & $\pm0.064$ & +0.020 & $-$0.094\\
20.4 & 0.00134 & 0.213 & $\pm$0.016 &  $\pm0.014$ & +0.018 & $-$0.040\\
 & 0.00588 & 0.126 & $\pm$0.010 & $\pm0.014$ & +0.010 & $-$0.042\\
\hline
 & 0.00356 & 0.257 & $\pm$0.046 & $\pm0.057$ & +0.020 & $-$0.084\\
112.0 & 0.00738 & 0.238 & $\pm$0.030 &  $\pm0.039$ & +0.015 & $-$0.041\\
 & 0.03230 & 0.086 & $\pm$0.020 & $\pm0.018$ & +0.001 & $-$0.026\\
\hline
\end{tabular}
\caption{The combined $F_{2}^{c\bar{c}}$ values from the production cross sections of $D^{\pm}$ and $D^{0}/\bar{D}^{0}$ not coming from $D^{*\pm}$ at each $Q^{2}$ and $x$ value. The statistical, systematic and extrapolation uncertainties are shown separately. The measurements have a further uncertainty of $3.3\%$ from the $D^{+} \rightarrow K^{-}\pi^{+}\pi^{+} (+c.c.)$ and  $D^{0} \rightarrow K^{-}\pi^{+} (+c.c.)$ branching ratios. The additional uncertainty from the luminosity measurement is $2.6\%$.}
\label{tab:f2cc_comb}
\end{center}
\end{table}

%
%
\end{document}